\documentclass[12pt,a4paper]{article}
\usepackage[pdfstartview=FitH,colorlinks=true,linkcolor=blue,anchorcolor=black,citecolor=black,urlcolor=blue]{hyperref}
\usepackage[english]{babel}
\usepackage{amsmath,amssymb,titling,authblk}
\usepackage{slashed}
\usepackage{amsmath}
\usepackage{amscd}
\usepackage[normalem]{ulem}
\usepackage{appendix}
\usepackage{bbold}
\usepackage{bm}

\usepackage{pdflscape}
\usepackage{yfonts}
\usepackage{amscd}
\usepackage{epsfig}
 \usepackage{microtype}

\usepackage{cite}

\allowdisplaybreaks[4]
\numberwithin{equation}{section}

\usepackage {color}
\usepackage{bbold}
\usepackage{braket}
\usepackage{bbm}

\definecolor{verde}{cmyk}{.83,.21,1,.08}

\advance\hoffset by -1.1truecm
\advance\voffset by -1.0truecm
\newcommand{\be}{\begin{equation}}
\newcommand{\ee}{\end{equation}}
\newcommand{\bea}{\begin{eqnarray}}
\newcommand{\eea}{\end{eqnarray}}

\newcommand{\del}{\partial}
\newcommand{\dd}{\mathrm d}
\newcommand{\ii}{\mathrm i}
\newcommand{\e}{\mathrm e}
\usepackage{mathrsfs}
\usepackage{authblk}

\numberwithin{equation}{section}
\usepackage{float}
\restylefloat{figure}
\newcounter{appendice}

\begin{document}

\setlength{\droptitle}{-6pc}

\title{Localization and Reference Frames\\in $\kappa$-Minkowski Spacetime}

\renewcommand\Affilfont{\itshape}
\setlength{\affilsep}{1.5em}

\author[1,2,3]{Fedele Lizzi\thanks{fedele.lizzi@na.infn.it}}
\author[1,2]{Mattia Manfredonia\thanks{mattia.manfredonia@na.infn.it}}
\author[4]{Flavio Mercati\thanks{flavio.mercati@gmail.com}}
\author[5]{Timoth{\'e}~Poulain\thanks{timothe.poulain@th.u-psud.fr}}
\affil[1]{Dipartimento di Fisica ``Ettore Pancini'', Universit\`{a} di Napoli {\sl Federico~II}, Napoli, Italy}
\affil[2]{INFN, Sezione di Napoli, Italy}
\affil[3]{Departament de F\'{\i}sica Qu\`antica i Astrof\'{\i}sica and Institut de C\'{\i}encies del Cosmos (ICCUB),
Universitat de Barcelona, Barcelona, Spain}
\affil[4]{Dipartimento  di  Fisica,  Sapienza  Universit\`a  di  Roma,
Roma,  Italy.}
\affil[5]{Laboratoire de Physique Th{\'e}orique, B\^at. 210,
CNRS and Universit\'e de Paris-Sud 11, Orsay, France}

\date{}

\maketitle

\vspace{-2cm}

\begin{abstract}\noindent
We study the limits to the localizability of events and reference frames in the $\kappa$-Minkowski quantum spacetime. Our main tool will be a representation of the $\kappa$-Minkowski commutation relations between coordinates, and the operator and measurement theory borrowed from ordinary quantum mechanics. Spacetime coordinates are described by operators on a Hilbert space, and a complete set of commuting observables cannot contain the radial coordinate and time at the same time. The transformation between the complete sets turns out to be the Mellin transform, which allows us to discuss the localizability properties of states both in space and time.
We then discuss the transformation rules between inertial observers, which are described by the quantum $\kappa$-Poincar\'e group. These too are subject to limitations in the localizability of states, which impose further restrictions on the ability of an observer to localize events defined in a different observer's reference frame.  
\end{abstract}

\newpage

\section{Introduction}

The problem of Quantum Gravity suggests that the classical spacetimes at the basis of General Relativity and Quantum Field Theory may have to be replaced with quantum structures. A concrete realization of this idea is provided by noncommutative geometry.
 In this paper we consider the $\kappa$-Minkowski space~\cite{kmink1,majid22,koso,koso2,Agostini2002,Dimitrijevic2003,
 MicheleKappa,FlavioKappaDifferential,FlavioKappaLightCone,TPJCW1,TPJCW2}, which is the homogeneous space of the $\kappa$-Poincar\'e Hopf algebra (quantum group)~\cite{LukierskiInventsKpoincare1,LukierskiInventsKpoincare2,
 ZakrzewskiInventsKPGroup,Lukierski_kappaPoincareanydimension,
 UniquenessOfKappa}. The commutation relations of the coordinate functions for $\kappa$-Minkowski are:
\be
[ x^0, x^i]=\ii\lambda  x^i, \ [ x^i, x^j]=0, \ i,j=1,2,3. \label{commrel}
\ee
Often the deformation parameter $\lambda$ is indicated by $\frac 1\kappa$, hence the name. For us, as usual $ x^0=c \,t$, $c$ being the speed of light, $\lambda$ has the dimension of a length, and a natural scale for time is given by $\frac\lambda c$. The coordinate operators are assumed Hermitian $(x^\mu)^\dagger = x^\mu$. 
Our aim is to study the geometrical kinematics of spacetime, seen as a ``quantum'' object.  The quantization parameter will be $\lambda$, a quantity presumably of the order of Planck length. Relations~\eqref{commrel} suggest we use the theory of operators on a Hilbert space as the correct description. Since we are interested in the kinematics of spacetime alone, and will not discuss momentum for $\kappa$-Minkowski, the quantum of action $\hbar$ will not play a role, except when we reason in analogy with particle quantum mechanics.

The geometry described by~\eqref{commrel} is a \emph{noncommutative geometry}. One of the aims of this paper is to discuss what sort of measurements of position and time are possible, and which are the states. Clearly the presence of nontrivial commutation relations indicates that a version of Heisenberg's uncertainty relations is present:
\be
\Delta x^0 \Delta x^i \geq \frac\lambda2 |\langle x^i \rangle| \,, \label{uncertx0xi}
\ee 
and it will not be possible in general to localise states both in space and in time. 
In our treatment we will follow Dirac's correspondence principle, \emph{i.e.} associate to the classical coordinates, and in general to the observables, operators on a Hilbert space, and consider their spectrum and eigenfunctions. We will also assume the eigenvalues to be the possible results of a measurement of the observables, and use the standard apparatus of quantum mechanics (although, we repeat, we do not consider conjugate momenta and their commutations).

Let us make more precise what we mean by noncommutative geometry. An ordinary topological space is fully described by the algebra of continuous complex-valued function (in the noncompact case, vanishing at infinity,) on it. These form a commutative $C^*$-algebra, which can always be represented as operators on a Hilbert space. Further structures, such as smoothness, are encoded in other operators such as the Dirac operator, or its generalizations (for a review see for example~\cite{WSS}). Usually one introduces a deformation of this algebra by defining a noncommutative deformed $\star$-product so that the $\star$-commutator $[ x^\mu, x^\nu]_\star= x^\mu\star  x^\nu- x^\nu\star  x^\mu$ reproduces~\eqref{commrel}, usually based on the composition of plane waves~\cite{Agostini2002, TPJCW3}. There exist many versions of $\star$-products which reproduce the commutatutation relation \eqref{commrel}, see \emph{e.g.} \cite{DurhuusSitarz, Meljanac:2008ud, TPJCW1}. One of them has proved useful for the study of the quantum properties of various models of $\kappa$-Poincar\'e invariant scalar field theories \cite{TPJCW1,TPJCW2}. Besides, the geometric (spectral) properties, \textit{\`a la} Connes, of the $\kappa$-Minkowski spacetime have been investigated in \cite{dandrea,marseille1,marseille2,Matassa}.

We are interested in the localizability of the states, \emph{i.e.} the possibility to have a state of the system which describes a pointlike event, or a good approximation of it. In a noncommutative geometry, such as the quantum phase space of a particle, it may not be possible to localize points due to some version of the uncertainty principle~(\ref{uncertx0xi}).  One might wonder whether the localizability properties of a state depend on the reference frame or not. This is not the case for the quantum phase space of one particle: the algebra of positions and momenta is invariant under classical translations and rotations. However, the algebra~\eqref{commrel} is clearly not invariant under the classical action of the Poincar\'e group (in particular under translations and boosts). It is however invariant under a noncommutative generalization of the Poincar\'e group - as a matter of fact, it is \emph{defined} as the homogeneous space of such generalization. This deformation of the Poincar\'e group makes the group manifold itself into a noncommutative space, and the transformation parameters relating different reference frames are subject to limitations to their localizability as well. As a consequence, different observers will not agree in general on the localizability properties of the same state.

\subsection*{On localization and pure states}\label{On_localization_and_pure_states}

Consider first the phase space of \emph{classical} mechanics, described through the commutative algebra of position and momentum operators $q$ and $p$. Probability distributions $\rho(p,q)$ are only required to be integrable, so they belong to the function space $L^1(\mathbbm{R}^{2d})$. We can represent the algebra as multiplication operators on $L^1(\mathbbm{R}^{2d})$, and bounded operators will be continuous functions which vanish at infinity. However a vector of $L^1(\mathbbm{R}^{2d})$, 
being a function, is not a pure state, because it can always be written as the sum of two vectors obtained, for example, by setting the original function to zero for $q^1>0$ or $q^1<0$, and adjusting the normalizations. Nevertheless there are pure states, which can be obtained as limits: the Dirac $\delta$'s, also called  evaluation maps in this case. So that if $f$ is a function the state is $\delta_{q_0,p_0}(f)=f(q_0,p_0)$.  This is true for all commutative algebras. These states correspond also to irreducible representations and can be used to reconstruct the topology. The $\delta$ is not a vector of $L^1(\mathbb R^{2d})$, but is an acceptable distribution, and can be reached as a limit of normalized vectors.

When the algebra is noncommutative, this kind of pure states does not usually exist. Think for example of the quantum-mechanical phase space algebra, \emph{i.e.} the algebra of bounded operators of $p$ and $q$, where $[p,q]=i \hbar$. In this case $\mathcal H$ is $L^2(\mathbb R^d)$, the space of wavefunctions, and pure states are any vector, while mixed states are mixed density matrices. The noncommutativity of the algebra implies that there are no states which correspond to a single localised phase space point. Pure states in this case are normalized vectors of $L^2(\mathbb R^d)$, the ``wave functions''.\footnote{Treating the quantum phase space as a noncommutative geometry, in 2 dimensions one gets the Moyal plane. It is easy to show that the $*$-product of a real function by itself is not definite positive, and therefore the evaluation maps cannot be states.}
This is of course a manifestation of Heisenberg uncertainty principle:
\begin{equation}
\Delta p \Delta q \geq \frac \hbar 2 \,,
\end{equation}
which forbids the localization of phase space regions of area smaller than $\frac \hbar 2$. 

In what follows we will be studying the states on the algebra~(\ref{commrel}), in a spirit similar to what described here in the case of classical and quantum mechanics. In particular we will focus on their localizability properties (\emph{i.e.} to what extent one can be certain that an event took place within a certain region of an observer's coordinate system), and on the relationship between the states measured by different inertial observers. To achieve this, we will make use of specific representations of the commutation relations~(\ref{commrel}) as operators acting on some Hilbert space of functions.

\subsection*{Outline of the paper}
In Sect.~\ref{se:statesandevents} we discuss the notions of states and events in a $\kappa$-Minkowski spacetime. To set the scene we first present the well-known case study of ordinary quantum phase space in Subsection~\ref{se:phasespace}. Proceeding in analogy we present the case of time and position in $\kappa$-Minkowski in Sect.~\ref{se:timeandpositioninkappa}, introducing the time operator and connecting its spectrum with Mellin transforms. We discuss the localisation of states for an observer at the origin. In Sect.~\ref{se:Poincare} we briefly introduce the $\kappa$-Poincar\'e symmetries of our space, and in Sect.~\ref{se:observers} we discuss the role of observer located away form the origin, using the deformed $\kappa$-Poincar\'e symmetry. This section is partly in 1+1 dimension, where explicit representations are easier to control. A consistent part of the section is devoted to the physical interpretation of the results. A final section contains conclusions and outlook.

\section{$\kappa$-Minkowski spacetime: states and events\label{se:statesandevents}}

In this section we present a discussion on the states of the algebra of $\kappa$-Minkowski spacetime. To set the scene however we first present the well know case of the single particle quantum phase space of ordinary quantum mechanics.

\subsection{A Case Study: the quantum phase space}\label{se:phasespace}

Before we consider $\kappa$-Minkowski space it is useful to consider the archetypical noncommutative geometry, that of the phase space of a single quantum particle. The content of this section is well known to every undergraduate student in physics, but we present it to set up a parallelism with what we will do in the next section.

A particle in three dimensions has a phase space which is a six-dimensional space spanned by the coordinates $( q^i, p_i)$. What makes the particle quantum is promoting these coordinates to operators $( \hat q^i, \hat p_i)$ with nonvanishing commutation relations
\be
[ \hat q^i, \hat p_j]=\ii\hbar\delta^i_j \ , \label{Heisencommrel}
\ee
all other commutators being zero. 
The most common representations of position and momenta is as operators acting on the Hilbert space of square integrable functions of position, $L^2(\mathbb R^3_q)$ as\footnote{To simplify the notation we indicate by $q$ and $p$ the corresponding three-vectors, avoiding the use of a notation like $\vec q$.}:
\be
\hat q^i\psi(q)=q^i\psi(q) \ ; \qquad \hat p_i\psi(q)=-\ii\hbar\frac\del{\del q^i} \psi(q) \ . \label{pqreponq}
\ee
We will indicate the operators with a hat $\hat~$.
Both the $\hat q$'s and $\hat p$'s are unbounded selfadjoint operators with a dense domain. The spectrum is the real line (for each $i$). They have no eigenvectors but have improper eigenfunctions, namely the eigenvalue problem is solved by a distribution. Since the $\hat q^i$'s commute among themselves it is possible to have a simultaneous improper eigenvector of all of them, these are the Dirac distributions $\delta(q-\bar q)$ for a particular position $\bar q$, which is a vector in $\mathbb R^3$. Similarly, for a particular momentum $\bar p$,  the improper eigenfunctions of the $\hat p_i$ are the plane waves $\e^{\ii \bar p_i q^i}$.

Formally, the eigenvalue equation
\be
\del_q \psi(q)= \alpha \psi(q) \ , \qquad \alpha \in \mathbbm{C}^3 \label{momeigenfline}
\ee 
is solved by any function of the kind $\e^{ \alpha \cdot q }$. No function of this kind is square integrable, and therefore there are no eigenvalues and (proper) eigenfunctions. The operator $\hat p$ is self-adjoint on the domain of absolutely continuous functions, which is dense in $L^2(\mathbb{R}^3_q)$. One can see from Eq.~\eqref{momeigenfline} that $\alpha$ must be pure imaginary, $\alpha=\ii k$, $k \in \mathbbm{R}^3$, for distributions to be well defined on the domain of selfadjointness of the operators.  If  $\alpha$ had a real part $e^{\alpha\cdot q}$ would not be a solution of the eigenvalue problem even in the distributional sense. The improper eigenfunctions of momentum are physically interpreted as infinite plane waves of precise frequency. Since plane waves are not vectors of the Hilbert space there is no quantum state which would give as measure exactly the value $\hbar k$, nevertheless we have all learned to live with this fact, and there is a well-defined sense in which we talk about ``particles of momentum~$\hbar k$''.

The representation~\eqref{pqreponq} is tantamount to the choice of $\hat q^i$ as a complete set of observables, and to the description of a quantum state as a function of positions. As usual we interpret  $|\psi(q)|^2$ for normalized functions as the  probability density to find the particle at position $q$. The wave function, being a complex quantity, contains also the information about the density probability of the momentum operator. The connection is in the choice of the complete set of commuting observables and Fourier transform. It is important that the Fourier transform is an isometry, \emph{i.e.} it maps normalized functions of positions into normalized functions of momenta. 

If we choose the $\hat p_i$ as the complete set, then it is natural to express the state of the system as a function of the $p$'s on which 
\be
\hat q^i\phi(p)=\ii\hbar \frac\del{\del p^i}\phi(p) \ ; \qquad \hat p_i\phi(p)=p_i \phi(p) \ . \label{pqreponp}
\ee
The functions $\psi(q)$ and $\phi(p)$ carry exactly the same information and are connected by a \emph{Fourier} transform, which is but an expansion on the eigenfunction of $\hat p$.
\be
\psi(q)=\frac 1{(2\pi)^{\frac32}} \int \dd^3 p\, \phi(p) \e^{\frac\ii\hbar p\cdot q}
\ee

The fact that $\hat p$ and $\hat q$ have been treated symmetrically (apart from signs) can be traced back to the symmetry of~\eqref{Heisencommrel}. If we choose a different set of commuting observables, for example the number operator, the total angular momentum and one of its components, the Hilbert space will look different (especially because these operators have discrete spectrum).

All this is of course well known. Let us now consider the case of $\kappa$-Minkowski in the same spirit.

\subsection{Time and position of events in $\kappa$-Minkowski \label{se:timeandpositioninkappa}}
In this section we will use the techniques of the previous section. Let us begin by considering the $\hat x^i$'s as a complete set of observables on the  Hilbert space $L^2(\mathbb R^3_x)$. We will represent the $\hat x^\mu$ as operators on this space.

\subsubsection{The operator representation}

The representations of the algebra generated by~\eqref{commrel} are discussed in detail in~\cite{Agostini:2005mf, DabrowkiPiacitelli}. In particular the paper of Dabrowski and Piacitelli has been an important inspiration. In the following, we focus on the representation of time and position operators given by
\bea\label{Rep_kappa-Minkowski}
\hat x^i \psi(x)&=&x^i\psi(x), \nonumber\\
\hat x^0\psi(x)&=&\ii \lambda \left(\sum_i x^i\del_{x^i} + \frac32\right)\psi(x)=\ii\lambda \left(r\del_r + \frac32\right)\psi(x).
\eea
The $\frac32$ factor is necessary to have symmetric operators. In $d$ dimensions 
$\frac12(r\del_r+\del_r r)=r\del_r+\frac d2$.
Here, $\hat x^0$ plays the role that $\hat p$ played in Sect.~\ref{se:phasespace}. 

The representation~\eqref{Rep_kappa-Minkowski} is far from being unique. In~\cite{MeljanacStojic} Meljanac and Stojic have written (in the Euclidean context) the most general class of operator with the correct characteristic, and shown that they depend on two functions with some constraints. It would be interesting to consider these more general realisations, but we will not do it in this paper (see also~\cite{MeljanacMercati,MeljanacMercati2}).

The $\hat x^0$ operator is, up to constants, the dilation operator, and this suggests the use of a polar basis. The polar coordinates $\hat \theta,\hat \varphi$ do not correspond to well defined self-adjoint operators, but we note that, defining $\hat r \cos \hat \theta=\hat x^3$ and $\hat  r \, \e^{\ii\hat \varphi}=(\hat x^1+\ii \hat x^2)$, a simple calculation shows that
\be
[\hat x^0, \cos \hat \theta]=[\hat x^0,\e^{\ii \hat \varphi}]= 0 \,, \qquad [\hat x^0,\hat r ] = \ii \lambda \hat r \,.
\ee
In fact $\hat x^0$ commutes with all spherical harmonics, or in general functions of $\hat \theta$ and $\hat \varphi$ independent on $r$.  Hence in the following we will consider the vectors of $L^2(\mathbb R^3_x)$ to be functions of the kind $\psi=\sum_{lm}\psi_{lm}(r)Y_{lm}(\theta,\varphi)$. Moreover, since the angular variables commute with everything, we will often concentrate on the radial parts, and consider functions of $r$ alone.
The uncertainty principle~\eqref{uncertx0xi} has its polar version
\be
\Delta \hat  x^0 \Delta \hat r \geq \frac\lambda2 |\langle \hat  r \rangle|. \label{uncertx0r}
\ee

The operator $\hat x^0$ is symmetric, but we should verify its self-adjointness domain. Since problems can only arise from the integration over $r$ we will assume that the angular degrees of freedom have been integrated out. Integrating by parts, one finds:
\be
\int \dd r  r^2\, \psi_1^* \ii \lambda\left(r\del_r + \frac32\right) \psi_2= \ii \lambda\int \dd r  r^2\, \psi_1^* \frac32\ \psi_2-\int\dd r\,  \ii\lambda \del_r\left( r^3 \psi_1^*\right) \psi_2 + \psi_1^* r^3 \psi_2 \bigg|^\infty_0 \,.
\ee
One can see that the boundary term vanishes if $\psi_1$ and $\psi_2$ vanish at infinity faster than  $r^{-\frac32}$, which is true for all square-integrable (according to the measure $\int\! \dd r r^2$) functions. In the origin the condition imposed is weaker than the one imposed by square-integrability.

Let us now look for the spectrum and the (improper) eigenvectors. They will be the equivalent of the plane waves. 
Monomial in $r$ are formal solutions of the eigenvalue problem:
\be
\ii\lambda \left(r\del_r + \frac32\right) r^\alpha=\ii\lambda (\alpha+\frac32) r^\alpha=\lambda_\alpha r^\alpha,
\ee
therefore eigenvalues are  
\be 
\lambda_\alpha= \ii \lambda(\alpha+\frac32).
\ee
These eigenvalues are real if and only if
\be
\alpha=-\frac32 + \ii \tau ,
\ee
with $-\infty<\tau<\infty$ a real number. In complete analogy with the momentum case previously discussed, unless the real part of $\alpha$ is -3/2, the improper eigenfunctions would not be acceptable distributions. The spectrum of the time operator is real and goes from minus infinity to plus infinity.


The distributions
\be
T_\tau=\frac{r^{-\frac32-\ii\tau}}{\lambda^{-\ii\tau}}=r^{-\frac32}\e^{-\ii\tau\log\left(\frac r\lambda\right)},
\ee
 are for time in classical $\kappa$-Minkowski space what plane waves are for momentum in quantum phase space. They are not physical states (vector of $L^2(\mathbbm{R}^3_x)$) because their behaviour at the origin and at infinity is bad, but ``just about'', an epsilon slower at the origin and faster at infinity would do, but then they would not be eigenfunction of $\hat x^0$.  They have a well defined inner product with every vector in the domain of $\hat x^0$. The distribution has the correct dimension of a length to the 3/2, the factor of $\lambda$ is there to avoid taking the logarithm of a dimensional quantity. Since $\lambda$ is a natural scale for the model, this choice is natural, but not unique.

\subsubsection{The spectrum of time and Mellin transforms}

Since $\hat x^0$ is a selfadjoint operator, it will have a complete basis. 
As we said what matters is only the radial coordinates, we will leave $\theta$ and $\varphi$ unchanged. We can therefore use in our set of complete observables either $r$ or $\tau$.

As noted earlier the completeness of the observables implies that any function of $r$ can be \emph{isometrically} expanded in terms of the $T_\tau$.
\be \label{expandPSI}
\psi(r,\theta,\varphi)=\frac1{\sqrt{2\pi}}\int_{-\infty}^{\infty}\dd\tau r^{-\frac32}\e^{-\ii\tau\log\left(\frac r\lambda\right)} \widetilde\psi(\tau,\theta,\varphi).
\ee
The integral above suggests $\psi(r,\theta,\varphi)$ to be some kind of integral transform of $\widetilde\psi(\tau,\theta,\varphi)$, the analog in this context of the Fourier transform.

It is in fact a \emph{Mellin} transform. Given a locally integrable function $f(x)$ with $x\in (0,\infty)$, the integral
\begin{equation} \label{MellinTransformDEF}
\mathcal{M}[f,s]=\frac{1}{\sqrt{2  \pi}}\int_0^\infty \dd x \,x^{s-1} f(x) =\mathcal{F}(s)
\end{equation} 
defines the \textit{Mellin transform} of $f$, when \eqref{MellinTransformDEF} converges.  
The integral in \eqref{MellinTransformDEF} converges for  $Re(s)\in(A,B)$ where $A$ and $B$ are real numbers such that
\begin{equation}
f(x)=\left \{ \begin{array}{rcl}O\left(x^{-A-\epsilon}\right) & \mbox{as}  & \chi \rightarrow 0_+ 
\\
O\left(e^{-B+\epsilon}\right) &  \mbox{as} &  \chi \rightarrow +\infty
\end{array} \right. \ , \ \forall \epsilon>0 \ , \ A<B.
\end{equation}
The interval $(A,B)$ is the so called  \textit{strip of analyticity} of $\mathcal{M}[f,s]$.
The inverse of the Mellin transform is\footnote{A more detailed discussion can be found in \cite{ParisKarminski}. }:
\begin{equation}\label{MellinTransformDEFINV}
\mathcal{M}^{-1}[\mathcal{F}(s), x]=\frac{1}{\ii \sqrt{2 \pi }}\int_{C-\ii\infty}^{C+\ii \infty}\dd s\, x^{-s} \mathcal{F}(s), \ A<C<B
\end{equation}

We require a transform which is an isometry between square integrable functions of $r$ with measure $\dd r r^2$ and functions of $\tau$. Therefore we define:
\begin{equation}
\psi(r,\theta,\varphi)=\frac1{\sqrt{2\pi}}\int_{-\infty}^{\infty}\dd\tau\, r^{-\frac32}\e^{-\ii\tau\log\left(\frac r\lambda\right)} \widetilde\psi(\tau,\theta,\varphi)=\mathcal{M}^{-1}\left[\widetilde\psi(\tau,\theta,\varphi), \ 
r \right],
\end{equation}
\begin{equation}
\widetilde \psi(\tau,\theta,\varphi)=\frac1{\sqrt{2\pi}}\int_0^\infty \dd r\, r^{\frac12}\e^{\ii\tau\log\left(\frac r\lambda\right)} \psi(r,\theta,\varphi)=\mathcal{M}\left[\psi(r,\theta,\varphi), \-\frac{3}{2}+\ii \tau \right].
\end{equation}
Thus, $\widetilde \psi$ is the Mellin transform of $\psi$ with $s=3/2+\ii\tau$. Hereafter we will often omit the explicit dependence on $\theta$ and $\varphi$ when there is no confusion. 
The above-defined transformations conserve the norms:
\be
\int_0^\infty\dd r r^2 |\psi(r)|^2=\int_{-\infty}^\infty\dd\tau |\widetilde\psi(\tau)|^2
\ee
Likewise there is a Parseval identity:
\be
\langle\psi_1|\psi_2\rangle=\int_0^\infty \dd r\, r^2 \overline \psi_1(r)\psi_2(r)=\int_{-\infty}^\infty \dd\tau\overline {\widetilde \psi_2}(\tau)\widetilde\psi_1(\tau)=\langle\widetilde\psi_1|\widetilde\psi_2\rangle
\ee
Assuming the usual measurement theory, we have that the average time measured by a particle in the state described by $\psi$ with spherical symmetry given by:
\be
\langle \hat x^0\rangle_\psi=4\pi \int r^2\dd r \overline \psi(r) \ii\lambda\left(r\del_r+\frac32\right) \psi(r) \label{x0average}
\ee
If $\psi$ is real it results $\langle \hat x^0\rangle_\psi=0$. In fact
\be
\begin{gathered}
\int r^3 \dd r \overline \psi(r) \del_r\psi(r) = r^3|\psi|^2\bigg|_0^\infty-\int r^3 \dd r \psi(r) \del_r\overline  \psi(r)-3\int r^2 \dd r  |\psi(r)|^2
\\
\Downarrow
\\
\psi = \overline  \psi ~~~ \Rightarrow ~~~  \int r^3 \dd r \overline \psi(r) \del_r\psi(r) = - \frac 32 \int r^2 \dd r  |\psi(r)|^2 \ ,
\end{gathered}
\ee
which implies that the two terms in~\eqref{x0average} cancel each other. Hence only complex valued functions will have a nonzero mean value for a measurement of time. One may note the analogy with quantum phase space, where real functions have a vanishing mean value of the momentum.The probability of measuring a given value of $\tau$  is given by $|\widetilde \psi(\tau)|^2$ for normalised functions. 

To get familiar with this representation let us give a few examples.
Consider the following state, localized on a shell of radius $r_0$:  $\psi(r)=\delta(r-r_0)/r_0^2$. Then
\be
\widetilde\psi(\tau)=\frac1{\sqrt{2\pi}} r_0^{-\frac32} \left(\frac{r_0}\lambda\right)^{\ii\tau}=\frac1{\sqrt{2\pi}} r_0^{-\frac32}\e^{\ii \tau\log \left(\frac{r_0}\lambda\right)} \,, \label{deltatransf}
\ee
and the probability $|\psi (\tau)|^2$ does not depend on $\tau$, which means that all values of time are equally probable, just like in quantum mechanics, where a localised particle has all values of momentum equally probable. Not surprisingly the function $\widetilde\psi(\tau)$ in \eqref{deltatransf} is not normalizable. We can regularize the delta function by approximating it with a constant function with support on a ``thick spherical shell'':
\be
\psi(r)=\left\{\begin{array}{lc}0& r<R_1\\ \sqrt{\frac3{4\pi(R_2^3-R_1^3)}}&R_1\leq r \leq R_2\\ 0 & R_2<r\end{array}\right.
\label{FirstWavefunc}
\ee
its Mellin transform is:
\be
\widetilde\psi(\tau)=\frac1{\sqrt{2\pi}}\sqrt{\frac3{4\pi(R_2^3-R_1^3)}}\left(\frac{R_2^{\frac32+\ii\tau}-R_1^{\frac32+\ii\tau}}{\lambda^{\ii\tau}}\right) \frac2{3+2\ii\tau} \,, \label{FirstWavefuncMellin}
\ee
with probability density:
\be
|\widetilde\psi(\tau)|^2=\frac3{8\pi^2(R_2^3-R_1^3)}\left[ R_2^3+R_1^3-2 R_1^{\frac32}R_2^{\frac32}\cos \left(\tau\log\frac{R_2}{R_1}\right)\right] \frac4{9+4\tau^2}\,,
\label{FirstWavefuncMellinProbDensity}
\ee
which is an even function, which explains why the average value of $\hat x^0$ vanishes. The probability density~(\ref{FirstWavefuncMellinProbDensity}) now is not constant: it is now peaked around $\tau=0$ and it decreases like $\tau^{-2}$ away from the origin.
In the limit $R_1\to R_2$ the Mellin transform~(\ref{FirstWavefuncMellin}) tends to (be proportional to) the Mellin transform of the delta function,~(\ref{deltatransf}).
 
It is useful to have an idea of the dimensional quantities involved. If we call $t$ the eigenvalue of the time operator $\frac{x^0}c$, then $\tau=t \frac c\lambda$. Note that $\frac c\lambda$ is a dimensional quantity. If we choose for $\lambda$ the Planck length then $\frac c\lambda\sim 2\cdot 10^{43}$~Hz. In other words if $t=1$\,s, then $\tau=2\cdot 10^{43}$, an extremely large number. If $t$ is of the order of Planck time, then $\tau\sim 1$.

\subsection{Localized states}

The aim of this section is to show that the localisation properties (in space and time) of a particle at the origin are different from those away from it. To this extent we will consider the Hilbert space vectors for particles in the two cases, and compare them, and their limit to a distribution. The relation~\eqref{uncertx0r} implies a generalised uncertainty principle which will limit the simultaneous localisabilty of a particle in space and time, we wish to see its explicit consequences for localised states. We have chosen to present the results of this section using concrete examples for clarity, we do not however have at present a general theory encopassing all possible states. This will have to wait for further work.

\subsubsection{Point localised at a finite distance from the origin}

\begin{figure}[t]\center
\includegraphics[width=0.5 \textwidth]{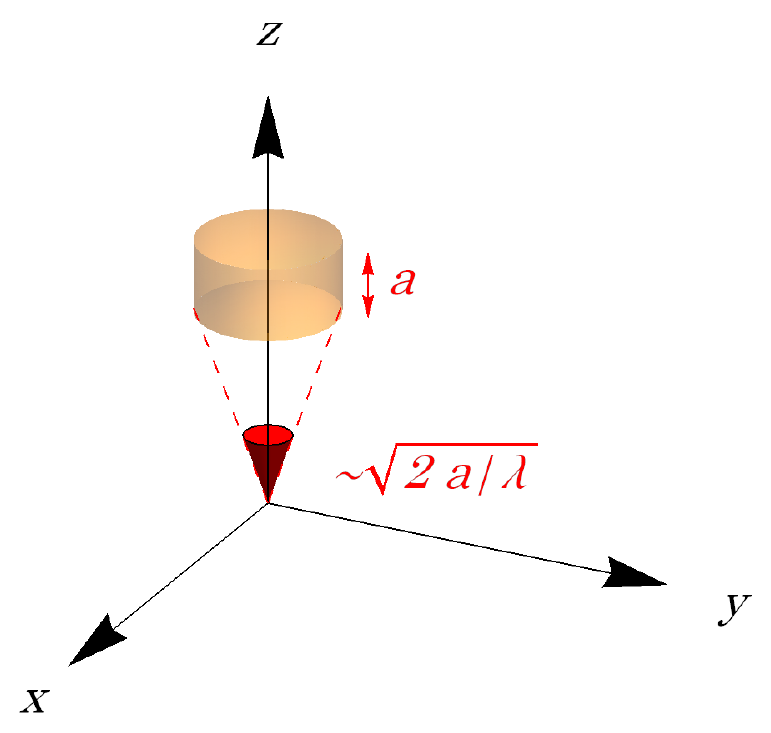}
\caption{\sl The support of the wavefunction~(\ref{ThirdWavefunc}).}
\label{Figure_Particle_localized_away_from_origin}
\end{figure}

\begin{figure}[t]\center
\includegraphics[width=0.7 \textwidth]{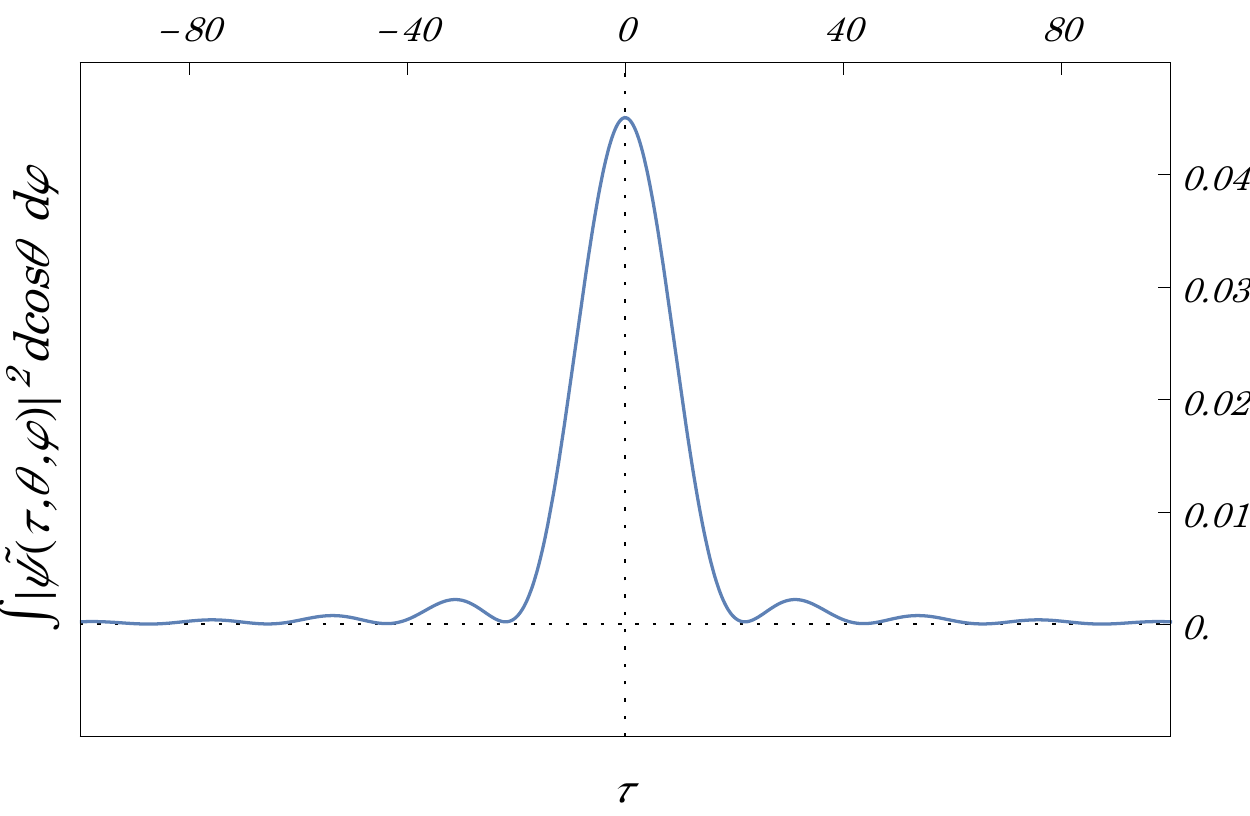}
\caption{\sl The $\tau$-dependence of the Mellin transform of the wavefunction~(\ref{ThirdWavefunc}).}
\end{figure}

Consider a wavefunction localised in space in a small region of size $a$ around a point at distance $z_0$ along the $z$ axis. The wavefunction can have constant value inside that region, and the normalization condition fixes that value. In spherical coordinate we can write:
\be
\psi_{z_0,a}(r,\theta,\varphi)=\left\{\begin{array}{ll}
\sqrt{\frac{3\lambda}{2 a \pi\left((a+z_0)^3-z_0^3\right)}} \,, & z_0\leq r\leq (z_0+a)\ \mbox{and}\ \cos\theta>1-\frac a\lambda\\
0, & \mbox{otherwise}
\end{array}\right.\label{ThirdWavefunc}
\ee
The shape of the region we are considering is shown in Fig.~\ref{Figure_Particle_localized_away_from_origin}. For any nonzero (positive) $a$ the wavefunction is normalized and is a well defined state of the Hilbert space $L^2(\mathbbm{R}^3_x)$. In the limit $a\to 0$ $\psi_{z_0,a}$ goes to a $\delta$ function localised at a distance $z_0$ from the origin along the positive $z$ axis. It is possible to calculate its Mellin transform:
\be
\widetilde\psi_{z_0}(\tau,\theta,\varphi)=
\frac{\sqrt{3\lambda}}{\pi }\ \   \frac{(z_0+a)^{\frac32+\ii\tau}-z_0^{\frac32+\ii\tau}}{\lambda^{\ii\tau}(3+\ii 2\tau)\sqrt{a \left((a+z_0)^3-z_0^3\right)}} \ 
\Theta\left(\cos \theta -1 +\frac a \lambda \right)
\ee
and the associated probability density:
\be
\begin{aligned}
|\widetilde\psi_{z_0,a}|^2 &=
\frac{3 \lambda}{\pi ^2} \frac{z_0^3+(z_0+a_0)^3-2 \left(z_0 \left(a+z_0\right)\right){}^{3/2} \cos \left(\tau  \log \left(\frac{z_0}{z_0+a}\right)\right)}{ \left(4 \tau ^2+9\right) a\left((a+z_0)^3-z_0^3\right)} \Theta\left(\cos \theta -1 +\frac a \lambda \right)
\\
&= \left[ \frac\lambda{4\pi^2 z_0} -\frac{\lambda a}{8 (\pi^2 z_0^2)} +\mathcal O (a^2) \right]  \Theta\left(\cos \theta -1 +\frac a \lambda \right)
\end{aligned}
\ee
We can integrate the above function in $\theta$, which gives a factor $a/\lambda$:
\bea
\int |\widetilde\psi_{z_0,a}|^2 \sin \theta \, \dd \theta =   \frac a {4\pi^2 z_0} -\frac{a^2}{8 \lambda (\pi^2 z_0^2)} +\mathcal O (a^3)
\eea
In the limit $a \to 0$, the Mellin-transformed wavefunction tends to a constant $\frac\lambda{4\pi^2 z_0}$ localized in $\theta$ in a cone of angle $\arccos(1- \frac a \lambda) - \pi/2 \sim \sqrt{ \frac {2a}{\lambda}}$. The angular average tends to a constant which vanishes as $a \to 0$ (because of the normalization). This 
implies that in the limit the state is not an $L^2$ function anymore, and is instead a function with zero scalar product with all $L^2$ functions.

Note also that (not surprisingly) the series expansion for $a$ around 0, and $z_0$ around $\infty$ are the same:
\be
\begin{aligned}
|\widetilde\psi_{z_0}|^2 &=
\frac{\lambda }{4 \pi ^2 z_0}-\frac{a \lambda }{8 \pi ^2
   z_0^2}+\frac{a^2 \lambda  \left(7-4 \tau ^2\right)}{192 \pi ^2
   z_0^3}+\mathrm O\left(a^3\right)\nonumber\\
   &=\frac{\lambda }{4 \pi ^2 z_0}-\frac{a \lambda }{8 \pi ^2
   z_0^2}+\frac{a^2 \lambda  \left(7-4 \tau ^2\right)}{192 \pi ^2
   z_0^3}+\mathrm O\left( z_0^{-4} \right)
\end{aligned}
\ee
This means that a sharp localization of a particle far away from the origin implies that the particle cannot be localised in time. And this is in accordance with the generalised uncertainty principle~\eqref{uncertx0r}.

\subsection{Points localized at the origin of space and limit to eigenstates of the origin}

\begin{figure}[h!]\center
\includegraphics[width=0.7 \textwidth]{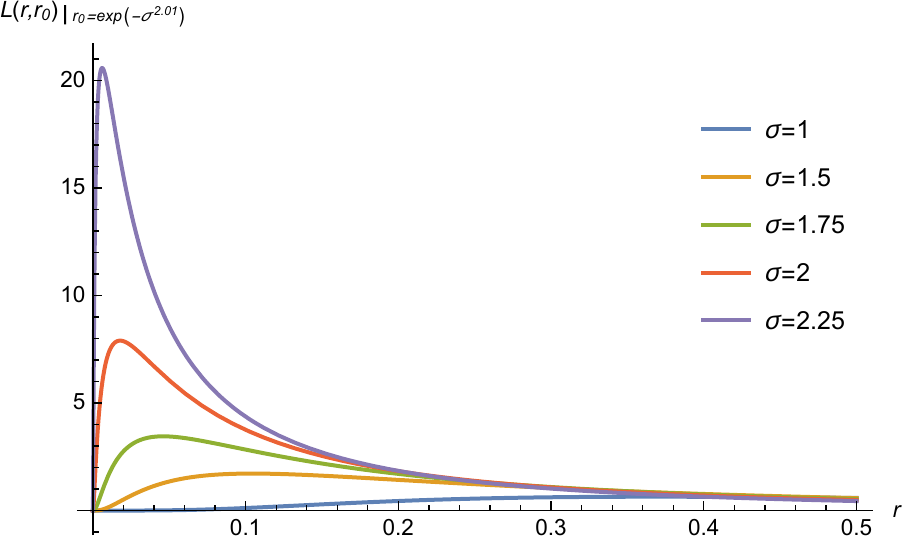}
\caption{\sl The $\sigma \to \infty $ limit of $L(r,r_0)$ when $\xi = e^{-\sigma^{(2+\epsilon)}}$, for $\epsilon = 0.01$. \label{figL}}
\end{figure}

We now present a one-parameter family of $L^2$ functions which tends to a state completely localized at the spatial origin (while in time it might be either completely localized around any value of $\tau$, or it may be nonlocal).
This is all allowed by the $\kappa$-Minkowski uncertainty relations~(\ref{uncertx0xi}), in which the presence of $\langle \hat x^i \rangle$ on the right hand side suggests that, although general localized states are impossible to achieve, in the special case of states localized at the spatial origin, perfect localization should be possible. Just like delta functions and plane waves in ordinary quantum mechanics (as described in Sec.~\ref{On_localization_and_pure_states}), it should be possible to obtain the mentioned states localized at the spatial origin as limits of normalized vectors of our Hilbert space. The key is to find functions that saturate the uncertainty bounds. In the case of the quantum phase space algebra, these are Gaussians (coherent states), as is well known. The $\kappa$-Minkowski algebra however is not canonical, and Gaussians are not minimal uncertainty states for this algebra. This role is played by \emph{log-Gaussians} normalized wavefunction, plotted in Fig.~\ref{figL},
\be\label{LogGaussians}
L(r,r_0)=N \e^{-\frac{(\log r - \log r_0)^2}{\sigma^2}}=\frac{\e^{-\left(\frac{\log
   \left(\frac{r}{r_0}\right)}\sigma\right)^2}\e^{-
   \frac{9}{16} \sigma^2}}{\sqrt{\sigma} (2 \pi
   )^{3/4} \sqrt{r_0^3}} \,.
\ee
They have  a maximum in $r=r_0$, and they localize at $r =r_0$ as $\sigma \to 0$, and  at $r=0$ as $r_0 \to 0$, for any value of  $\sigma \geq 0$.

The calculation of the average values of $\hat r^n$ is straightforward:
\be\label{ExpectationValue_r^n}
\langle \hat r^n \rangle_L=\e^{\frac{\sigma^2}8 n(n+6)} r_0^n \,,
\ee
and shows that they all vanish for $r_0 \to 0$.
In order to calculate the quantity $\langle r^n \rangle_L$ it is best to Mellin transform, the function in $\tau$ space is remarkably simple:
\be
\widetilde L(\tau,r_0)=\frac{\sigma^{\frac12}\e^{-\frac{1}{4} \sigma^2\tau 
   (\tau -3 \ii)}  }{2
   \sqrt[4]{2} \pi ^{3/4}}  \left(\frac{r_0}{\lambda}\right)^{\ii \tau } \,,
\ee
The interesting fact is that:
\be
\left|\widetilde L(\tau,r_0)\right|^2=\frac{\sigma\e^{-\frac{\sigma^2 \tau
   ^2}{2}}}{4 \sqrt{2} \pi
   ^{3/2}} \,,
\ee
namely, in $\tau$ space the probability density is a Gaussian independent on $r_0$. It is now trivial to see that
\be\label{ExpectationValue_x0^n}
 \langle (\hat x^0)^n \rangle_L = \frac 1 {4 \pi} \left(\frac\lambda\sigma\right)^n \left\{ \begin{array}{lcl} 0 & \ & n\ \mbox{odd}\\
(n-1)!! & \ & n\ \mbox{even}\end{array}\right.
\ee
We can see that there is a double limit $r_0\to 0$ and $\sigma\to\infty$\footnote{For example, it is sufficient to take $r_0=e^{-\sigma^{2+\epsilon}}$ for any $\epsilon>0$, that all $\langle \hat r^n \rangle_L$ in~(\ref{ExpectationValue_r^n}) and all $\langle (\hat x^0)^n \rangle_L$ in~(\ref{ExpectationValue_x0^n}) go to zero as $\sigma \to \infty$.} which gives a state which is localised both is space (at $r=0$) and in time. In the example above the time localization is at $\tau=0$, but it is possible to shift the state by multiplying the function by $r^{\ii\tau_0}$. Moreover one can attribute any wavefunction to time while still having the spatial coordinates localized at the origin, just by convoluting this with a function of $\tau$.
We have then introduced a state - which we can call `eigenstate of the origin', and refer to as\footnote{While we have seen that there is a state corresponding to $\ket o$, there is not a normalized vector corresponding to it. Here (and in the following) we are here performing the usual abuse of notation made when one uses the ket notation $\ket x$ in ordinary quantum mechanics. } $|o\rangle$ - that is completely localized at the origin of spacetime and can be obtained as a limit of normalized elements of $L^2(\mathbbm{R}^3_x)$. Moreover we have a 1-parameter family of states, which we indicate from now on with $|o_\tau\rangle$, which are localized at the origin of space, at a nonzero time. These states too can be obtained as limits of normalized elements of $L^2(\mathbbm{R}^3_x)$.

\section{$\kappa$-Poincar\'e Symmetry \label{se:Poincare}}

In this section we briefly introduce the deformed symmetry of our space. we still concentrate on the group rather than the algebra. We will opt for an intuitive presentation, rather than a mathematically rigorous one. In the following, to lighten the notation we will suppress the hat symbol we used so far to distinguish quantum operators.

\subsection{The $\kappa$-Poincar\'e quantum group}

The algebra~(\ref{commrel}) emerges as the quantum homogeneous space of a Hopf-algebra deformation of the Poincar\'e group, known as \emph{$\kappa$-Poincar\'e}~\cite{LukierskiInventsKpoincare1,LukierskiInventsKpoincare2,
ZakrzewskiInventsKPGroup,kmink1}. This object has historical precedence over $\kappa$-Minkowski, which was introduced by Majid and Ruegg after recognizing the `bicrossproduct' structure of 
the $\kappa$-Poincar\'e group~\cite{kmink1}. The $\kappa$-Poincar\'e group is part of a very small family of possible Hopf-algebra deformations of the Poincar\'e group with a deformation parameter with the dimensions of (the inverse of) an energy~\cite{ZakrzewskiPoissonStructures,UniquenessOfKappa}. Moreover, under the requirement of undeformed spatial isotropy, the version of $\kappa$-Poincar\'e corresponding to~(\ref{commrel}) is singled out uniquely~\cite{UniquenessOfKappa}.

We introduce $\kappa$-Poincar\'e as the noncommutative algebra of functions $\mathcal{P_\kappa}$, generated by $\Lambda^\mu{}_\nu$ and $a^\mu$ that leave the commutation relations~(\ref{commrel}) invariant under the transformation:
\begin{equation}\label{TransformedObserver}
x^\mu \to  x'^\mu  = \Lambda^\mu{}_\nu \otimes x^\nu + a^\mu \otimes 1 \,.
\end{equation}
We ask that the above map, from the $\kappa$-Minkowski algebra $\mathcal M_\kappa$ to the tensor product algebra $\mathcal{P_\kappa} \otimes \mathcal M_\kappa$, is a \emph{left-coaction}. This entails that the map is a homomorphism with respect to the noncommutative product of $\mathcal M_\kappa$, hence the covariance of the commutation relations~(\ref{commrel}). In other words, we require that
\begin{equation}
[ x'^\mu , x'^\nu ] =  \ii\lambda \left( \delta^\mu{}_0 \, x'^\nu - \delta^\nu{}_0 \, x'^\mu \right) \,,
\end{equation}
this fixes some commutation relations between the $\kappa$-Poincar\'e group coordinates\footnote{The metric used here is $\eta^{\mu\nu} = \text{diag}(+,-,-,-)$.}:
\begin{equation}\label{kappaPoincareGroup}
\begin{gathered}
\left[ a^\mu , a^\nu \right] =  \ii\lambda \left( \delta^\mu{}_0 \, a^\nu - \delta^\nu{}_0 \, a^\mu \right) \,, \qquad [ \Lambda^\mu{}_\nu ,  \Lambda^\rho{}_\sigma ] = 0 \,,
\\
[ \Lambda^\mu{}_\nu , a^\rho ] =  \ii\lambda \left[  \left( \Lambda^\mu{}_\sigma \delta^\sigma{}_0 - \delta^\mu{}_0 \right) \Lambda^\rho{}_\nu + \left( \Lambda^\sigma{}_\nu \delta^0{}_\sigma - \delta^0{}_\nu \right) \eta^{\mu\rho} \right] \,.
\end{gathered}
\end{equation}
Also, the group laws (the group product, or composition law, the inverse and the identity), here encoded with a \emph{coproduct} $\Delta : \mathcal{P_\kappa} \to \mathcal{P_\kappa} \otimes \mathcal{P_\kappa}$:
\begin{equation}
\Delta(a^\mu) = a^\nu \otimes \Lambda^\mu{}_\nu +1 \otimes a^\mu \,,
\qquad
\Delta ( \Lambda^\mu{}_\nu ) = \Lambda^\mu{}_\rho \otimes \Lambda^\rho{}_\nu \,,
\end{equation}
an \emph{antipode} $S : \mathcal{P_\kappa} \to \mathcal{P_\kappa}$
\begin{equation}
S(a^\mu) = - a^\nu  (\Lambda^{-1})^\mu{}_\nu \,,
\qquad
S ( \Lambda^\mu{}_\nu ) = (\Lambda^{-1})^\mu{}_\nu \,,
\end{equation}
and a \emph{counit} $\varepsilon : \mathcal{P_\kappa} \to \mathbbm{C}$,
\begin{equation}
\varepsilon(a^\mu) = 0 \,,
\qquad
\varepsilon ( \Lambda^\mu{}_\nu ) = \delta^\mu{}_\nu \,,
\end{equation}
have to be homorphisms with respect to the commutation relations~(\ref{kappaPoincareGroup}).  In this way we make sure that our noncommutative algebra of functions on the Poincar\'e group is compatible with the group structure.
Finally, in order to have a proper Hopf algebra, the group maps, together with the noncommutative product, have to satisfy two identities. One is the \emph{coassociativity} of the coproduct:
\begin{equation}
(\Delta \otimes \text{id} ) \circ \Delta = ( \text{id} \otimes \Delta ) \circ \Delta \,,
\end{equation}
which ensures that we can combine two coproducts in either order, and the result is the same. As we said, the coproduct encodes the group combination law. Combining two coproducts means that we are making three subsequent transformations in the two possible orders, and the combined transformation is the same. This is just one of the axioms of ordinary Lie groups: the associativity of the group product. The other axioms are the existence of the identity (ensured by the existence of the counit map), and the relation between the group inverse and the identity. This is now encoded in the \emph{Hopf identity:}
\begin{equation} \label{counitantipode}
\mu \circ ( S \otimes \text{id} ) \circ \Delta = 
\mu \circ ( \text{id} \otimes  S ) \circ \Delta  = \varepsilon \,,
\end{equation}
which ensures that the antipode provides a left- and right-inverse for the coproduct ($\mu :  \mathcal{P_\kappa} \otimes  \mathcal{P_\kappa} \to \mathcal{P_\kappa}$ stands for the (noncommutative) multiplication map).

\subsection{A representation of the $\kappa$-Poincar\'e quantum group}

The operators $\Lambda^\mu{}_\nu$ in~(\ref{kappaPoincareGroup}) should not be understood as 16 independent operators, but rather as 16 redundant functions satisfying the relations $\eta_{\mu\nu} \Lambda^\mu{}_\rho \Lambda^\nu{}_\sigma = \eta_{\rho\sigma}$,  which reduce the independent components to 6. Since all components of $\Lambda^\mu{}_\nu$ commute with each other, the standard representation theory of the Lorentz group applies, and we can write 
\begin{equation}
\Lambda^\mu{}_\nu = (\exp \omega)^\mu{}_\nu \,, \qquad  \omega^\mu{}_\rho \eta^{\rho\nu}= - \omega^\nu{}_\rho \eta^{\rho\mu} \,,
\end{equation}
where the (Lorentzian) antisymmetry relation above reduces the independent components of $\omega^\mu{}_\nu$ to 6. These components commute with each other:
\begin{equation}
[\omega^\mu{}_\nu , \omega^\rho{}_\sigma ] = 0\,,
\end{equation}
but  do \emph{not} commute with $a^\mu$. The structure of the commutation relations~(\ref{kappaPoincareGroup}) suggests to represent the $a^\mu$'s as vector fields:
\begin{equation}
a^\rho = -  \ii\lambda \left[  \left( \Lambda^\mu{}_\sigma \delta^\sigma{}_0 - \delta^\mu{}_0 \right) \Lambda^\rho{}_\nu + \left( \Lambda^\sigma{}_\nu \delta^0{}_\sigma - \delta^0{}_\nu \right) \eta^{\mu\rho} \right]  \frac{\partial}{\partial \Lambda^\mu{}_\nu}  \,,
\end{equation}
and the exponential relation between $\omega^\mu{}_\nu$ and $\Lambda^\mu{}_\nu$ implies $\frac{\partial}{\partial \Lambda^\mu{}_\nu}  =  \Lambda^\nu{}_\alpha  \frac{\partial}{\partial \omega^\mu{}_\alpha} $, which allows us to write the above representation as vector fields acting on the space of $\omega^\mu{}_\nu$ coordinates: 
\begin{equation}
a^\rho = - i \,  \lambda \left[  \left( \Lambda^\mu{}_\sigma \delta^\sigma{}_0 - \delta^\mu{}_0 \right) \Lambda^\rho{}_\nu + \left( \Lambda^\sigma{}_\nu \delta^0{}_\sigma - \delta^0{}_\nu \right) \eta^{\mu\rho} \right] \Lambda^\nu{}_\alpha  \frac{\partial}{\partial \omega^\mu{}_\alpha}  \,.
\end{equation}
Interestingly, the above vector fields already `know' about the commutation relations between the translation operators. In fact, the commutator of two of these vector fields acts on wavefunctions of $\omega^\mu{}_\nu$ as the Lie bracket between the vector fields, and computing this Lie bracket yields $\left[ a^\mu , a^\nu \right] =  \ii\lambda \left( \delta^\mu{}_0 \, a^\nu - \delta^\nu{}_0 \, a^\mu \right)$.

We found a representation of the $\kappa$-Poincar\'e algebra, in which $\Lambda^\mu{}_\nu$ represent as multiplication operators on wavefunctions of $\omega^\mu{}_\nu$:
\begin{equation}\label{ReprLorentzMatrices}
\Lambda^\mu{}_\nu \phi(\omega) = (\exp \omega)^\mu{}_\nu \phi(\omega) \,,
\end{equation}
while the translation operators act as vector fields:
\begin{equation}\label{Representation1_translations}
a^\rho \phi(\omega) = - i \,  \lambda \left[  \left( \Lambda^\mu{}_\sigma \delta^\sigma{}_0 - \delta^\mu{}_0 \right) \Lambda^\rho{}_\nu + \left( \Lambda^\sigma{}_\nu \delta^0{}_\sigma - \delta^0{}_\nu \right) \eta^{\mu\rho} \right] \Lambda^\nu{}_\alpha  \frac{\partial \phi(\omega)}{\partial \omega^\mu{}_\alpha}  \,.
\end{equation}
The wavefunctions can be taken as belonging to $L^2(SO(3,1))$, with the scalar product constructed, \emph{e.g.} with the Haar measure on the Lorentz group.

Unfortunately the representation we just considered is not good enough: it is not \emph{faithful}. In fact we can write combinations of the $\Lambda^\mu{}_\nu$ and $a^\rho$ operators that are represented into the null operator:
\begin{equation}
\begin{aligned}
&\eta_{\rho\mu} \left( \Lambda^\mu{}_\sigma \delta^\sigma{}_0 - \delta^\mu{}_0 \right) \, a^\rho \triangleright \phi(\omega) =
\\
&\left[\eta _{\rho \beta } \Lambda ^{\rho }{}_{\nu } \left(\delta ^{\kappa }{}_0 \Lambda ^{\beta }{}_{\kappa }-\delta ^{\beta }{}_0\right) \left(\delta ^{\sigma }{}_0 \Lambda ^{\mu }{}_{\sigma }-\delta ^{\mu }{}_0\right)+\left(\delta ^{\sigma }{}_0 \Lambda ^{\mu }{}_{\sigma }-\delta ^{\mu }{}_0\right) \left(\delta ^0{}_{\sigma } \Lambda ^{\sigma }{}_{\nu }-\delta ^0{}_{\nu }\right)\right] \Lambda^\nu{}_\alpha  \frac{\partial \phi(\omega)}{\partial \omega^\mu{}_\alpha} =
\\ 
&\left(\delta ^{\sigma }{}_0 \Lambda ^{\mu }{}_{\sigma }-\delta ^{\mu }{}_0\right)\left[\eta _{\rho \beta } \Lambda ^{\rho }{}_{\nu } \left(\delta ^{\kappa }{}_0 \Lambda ^{\beta }{}_{\kappa }-\delta ^{\beta }{}_0\right)+\left(\delta ^0{}_{\sigma } \Lambda ^{\sigma }{}_{\nu }-\delta ^0{}_{\nu }\right)\right] \Lambda^\nu{}_\alpha  \frac{\partial \phi(\omega)}{\partial \omega^\mu{}_\alpha}
=
\\
&\left(\delta ^{\sigma }{}_0 \Lambda ^{\mu }{}_{\sigma }-\delta ^{\mu }{}_0\right)\left[\left(\eta _{00}-1\right) \delta ^0{}_{\nu }+\left(1-\eta _{00}\right) \Lambda ^0{}_{\nu }\right]
 \Lambda^\nu{}_\alpha  \frac{\partial \phi(\omega)}{\partial \omega^\mu{}_\alpha} =0
\end{aligned}
\end{equation}
where the last line is zero because $\eta_{00} =+1$ in our convention. The operator 
\begin{equation}\label{Operator_that_represents_to_zero}
\eta_{\rho\mu} \left( \Lambda^\mu{}_\sigma \delta^\sigma{}_0 - \delta^\mu{}_0 \right) \, a^\rho \,,
\end{equation}
is nontrivial and, at least in order to admit a good classical limit, some of its expectation values should not be vanishing.
We conclude that the representation~(\ref{Representation1_translations}) is not faithful, and it needs to be enlarged. The simplest way to do it is to write a direct sum of representations: the above one and the  (at this point familiar) representation~(\ref{Rep_kappa-Minkowski}) of $\kappa$-Minkowski coordinates, which reproduces the commutation rules between translation operators, but commutes with Lorentz transformations. The Hilbert space now has to be enlarged with  three additional coordinates $q^i \in \mathbbm{R}$, $i=1,2,3$, 
so it is $L^2(SO(3,1)\times \mathbbm{R}^3)$, the Lorentz matrices still represent as multiplicative operators~(\ref{ReprLorentzMatrices}), and the translation operators are represented as follows:
\bea
a^\rho &=&   - \ii \,  \frac\lambda2 \left[  \left( \Lambda^\mu{}_\sigma \delta^\sigma{}_0 - \delta^\mu{}_0 \right) \Lambda^\rho{}_\nu + \left( \Lambda^\sigma{}_\nu \delta^0{}_\sigma - \delta^0{}_\nu \right) \eta^{\mu\rho} \right] \Lambda^\nu{}_\alpha  \frac{\partial}{\partial \omega^\mu{}_\alpha} \nonumber\\
&&+\ii\frac\lambda2 \left( \delta^\rho{}_0 \, q^i \frac{\partial}{\partial q^i} + \delta^\mu{}_i \, q^i \right) + \frac12\mbox{h.c.}
\eea
Where by ``h.c.'' we mean the hermitean conjugate of the previous expression. This ensure that the operator is self-adjoint on some domain.
The final form of our representation is
\begin{equation}
\begin{aligned}
a^\rho \phi(q,\omega)=&  \ii\lambda  \delta^\rho{}_0 \, \left( \frac 3 2 \, \phi(q,\omega)+ q^i \frac{\partial \phi(q,\omega)}{\partial q^i} \right)  + \delta^\mu{}_i \, q^i \, \phi(q,\omega) 
\\
& -  \ii\lambda: \left[ \left( \Lambda^\mu{}_\sigma \delta^\sigma{}_0 - \delta^\mu{}_0 \right) \Lambda^\rho{}_\nu + \left( \Lambda^\sigma{}_\nu \delta^0{}_\sigma - \delta^0{}_\nu \right) \eta^{\mu\rho} \right] \Lambda^\nu{}_\alpha  \frac{\partial}{\partial \omega^\mu{}_\alpha} : \, \phi(q,\omega) \,,
\\
\Lambda^\mu{}_\nu \phi(q,\omega) =& \Lambda^\mu{}_\nu(\omega) \phi(\omega)=(\exp \omega)^\mu{}_\nu \phi(q,\omega) \,,
\end{aligned}
\end{equation}
that is,
\begin{equation}\label{3+1kappaPoincareRepresentation}
\begin{aligned}
a^\rho \phi(q,\omega)=&  \ii\lambda  \delta^\rho{}_0 \, \left( \frac 3 2 \, \phi(q,\omega)+ q^i \frac{\partial \phi(q,\omega)}{\partial q^i} \right)  + \delta^\mu{}_i \, q^i \, \phi(q,\omega) 
\\
& - \frac { \ii\lambda} 2 \left[ \left( \Lambda^\mu{}_\sigma \delta^\sigma{}_0 - \delta^\mu{}_0 \right) \Lambda^\rho{}_\nu + \left( \Lambda^\sigma{}_\nu \delta^0{}_\sigma - \delta^0{}_\nu \right) \eta^{\mu\rho} \right] \Lambda^\nu{}_\alpha  \frac{\partial \phi(q,\omega)}{\partial \omega^\mu{}_\alpha} 
\\
& -\frac { \ii\lambda} 2  \phi(q,\omega)   \frac{\partial}{\partial \Lambda^\mu{}_\nu}\left[ \left( \Lambda^\mu{}_\sigma \delta^\sigma{}_0 - \delta^\mu{}_0 \right) \Lambda^\rho{}_\nu + \left( \Lambda^\sigma{}_\nu \delta^0{}_\sigma - \delta^0{}_\nu \right) \eta^{\mu\rho} \right] \,,
\\
\Lambda^\mu{}_\nu \phi(q,\omega) =& \Lambda^\mu{}_\nu(\omega) \phi(\omega)=(\exp \omega)^\mu{}_\nu \phi(q,\omega) \,.
\end{aligned}
\end{equation}

It is trivial to check that, since the derivatives with respect to $\omega^\mu{}_\nu$ commute with the functions of $q^i$,  and the derivatives with respect to $q^i$ commute with the functions of $\omega^\mu{}_\nu$, the representation splits into a direct sum of representations, and the commutation relations between $a^\mu$'s are satisfied.

The representation~\eqref{3+1kappaPoincareRepresentation} is complicated, and its explicit functional form depends on the  coordinate system on the Lorentz group we choose. In two spacetime dimensions the situation is greatly simplified by the fact that the Lorentz group is 1-dimensional, and everything can be made very explicit. In the next Subsection we will repeat the steps that led us to introduce the representation~(\ref{3+1kappaPoincareRepresentation}) in the 1+1-dimensional case, a useful exercise both for pedagogical reasons, and in order to have an example that can be worked out explicitly. This will be useful later.

\subsection{The representation of $\kappa$-Poincar\'e in 1+1 dimensions}

The great advantage of working in 1+1 dimensions is that we have an explicit (and simple) coordinatization of the Lorentz group:
\begin{equation}
\Lambda^0{}_0 = \Lambda^1{}_1 = \cosh \xi \,, 
\qquad
\Lambda^0{}_1 = \Lambda^1{}_0 = \sinh \xi \,, 
\end{equation}
in this parametrization. The commutation relations of  $\kappa$-Poincar\'e~\eqref{kappaPoincareGroup}
take the form
\begin{equation}\label{1+1_kappaPoincareGroup_v2}
\begin{gathered}
{}[ a^0 , a^1 ] =  \ii\lambda \, a^1  \,,
\qquad
[ \cosh \xi , a^0 ] = - \ii\lambda \sinh^2 \xi \,,
\qquad
[ \cosh \xi , a^1 ] = -  \ii\lambda  \left( \cosh \xi  - 1 \right)\sinh \xi  \,,
\\
[ \sinh \xi , a^0 ] = - \ii\lambda \sinh \xi  \cosh \xi \,,
\qquad
[ \sinh \xi , a^1 ] = -  \ii\lambda  \left( \cosh \xi  - 1 \right)\cosh \xi  \,,
\end{gathered}
\end{equation}
which can be simplified to:
\begin{equation}\label{1+1_kappaPoincareGroup_v3}
\begin{gathered}
{}[ a^0 , a^1 ] =  \ii\lambda \, a^1  \,,
\qquad
[  \xi , a^0 ] = - \ii\lambda \sinh \xi \,,
\qquad
[ \xi , a^1 ] =  \ii\lambda  \left( 1- \cosh \xi   \right) \,.
\end{gathered}
\end{equation}
It is evident that $a^0$ and $a^1$ act on $\xi$ like vector fields:
\begin{equation}
\begin{gathered}
a^0  =   \ii\lambda \sinh \xi \, \frac{\partial}{\partial \xi} \,,
\qquad
 a^1 =   \ii\lambda  \left( \cosh \xi  - 1 \right) \, \frac{\partial}{\partial \xi} \,.
\end{gathered}
\end{equation}
The above representation would be acceptable, as it reproduces the $[ a^0 , a^1 ]$ commutation relations. In this case we can easily show this explicitly:
\begin{equation}
\begin{aligned}
\left[a^0 ,a^1\right] &=- \lambda^2 \left[ \sinh \xi \, \frac{\partial}{\partial \xi} \left( \cosh \xi  - 1 \right) 
- \left( \cosh \xi  - 1 \right) \, \frac{\partial}{\partial \xi}
\sinh \xi  \right]  \frac{\partial}{\partial \xi}
\\
&=-\lambda^2 \left[ \sinh^2 \xi  
- \left( \cosh \xi  - 1 \right) \cosh \xi  \right]  \frac{\partial}{\partial \xi}
\\
&=-\lambda^2 \left( \cosh \xi  - 1 \right)  \frac{\partial}{\partial \xi} =  \ii \lambda a^1  \,.
\end{aligned}
\end{equation}
As before, this representation cannot be faithful, because the operator:
\begin{equation}
\left( \cosh \xi  - 1 \right) a^0 - \sinh \xi \, a^1  = -  \ii\lambda \left( \cosh \xi  - 1 \right) \sinh \xi \, \frac{\partial}{\partial \xi}+   \ii\lambda \sinh \xi  \left( \cosh \xi  - 1 \right) \, \frac{\partial}{\partial \xi} = 0 \,,
\end{equation}
which is the 1+1-dimensional version of~(\ref{Operator_that_represents_to_zero}), is represented into the null operator.
Again, it is sufficient to add to the above representation the familiar representation of the $\kappa$-Minkowski algebra in 1+1 dimensions:
\begin{equation}
\begin{gathered}
a^0  =  \ii\lambda q \frac{\partial}{\partial q}  +  \ii\lambda \sinh \xi \, \frac{\partial}{\partial \xi} \,,
\qquad
 a^1 = q +  \ii\lambda  \left( \cosh \xi  - 1 \right) \, \frac{\partial}{\partial \xi} \,,
\end{gathered}
\end{equation}
the two parts commute with each other, and separately satisfy the commutation relations and the Jacobi identity, and therefore they provide a good representation of our algebra on the Hilbert space $L^2(SO(1,1) \times \mathbbm{R}) \sim L^2(\mathbbm{R}^2)$ of square-integrable functions of $\xi$ and~$q$.
This representation is not selfadjoint, but it can be made so by Weyl-ordering it:

\bea
a^0 &=& \frac{\ii\lambda}2 \left( q \frac{\partial}{\partial q} +  \frac{\partial}{\partial q} q\right)  + {\frac{\ii\lambda}2} \left( \sinh \xi \, \frac{\partial}{\partial \xi} +  \frac{\partial}{\partial \xi} \sinh \xi \right) \nonumber\\
 a^1 &=& q + \frac{\ii\lambda}2 \left(  \left( \cosh \xi  - 1 \right) \, \frac{\partial}{\partial \xi} + \frac{\partial}{\partial \xi} \left( \cosh \xi  - 1 \right)  \right)  \,,
\eea
which can be written
\bea\label{final_1+1_k-Poinc_representation}
a^0  &=&  \ii\lambda \left(\frac 1 2 + q \frac{\partial}{\partial q} \right)  + \ii\lambda \left(  \frac 1 2 \cosh \xi  + \sinh \xi \, \frac{\partial}{\partial \xi}\right) \nonumber\\
 a^1 &=& q + \ii\lambda \left( \frac 1 2 \sinh \xi +  \left( \cosh \xi  - 1 \right) \, \frac{\partial}{\partial \xi}   \right)  \,.
\eea
It is easy to check that the above reproduces the commutation relations~(\ref{kappaPoincareGroup}).

\subsection{From $\kappa$-Poincar\'e to $\kappa$-Minkowski}

We can now make precise, within the framework of the representations we introduced for  $\kappa$-Minkowski and $\kappa$-Poincar\'e, in which sense the $\kappa$-Minkowski noncommutative spacetime is the quantum homogeneous space obtained by quotienting the $\kappa$-Poincar\'e quantum group by the Lorentz group.
The idea is that there are enough states in the representation of $\kappa$-Poincar\'e that we can reproduce any vector in the Hilbert space of the representation of $\kappa$-Minkowski [\emph{i.e.} $L^2(\mathbbm{R})$] as an appropriate limit of vectors belonging to the representation of $\kappa$-Poincar\'e  [$L^2(SO(3,1)\times \mathbbm{R})$], in which the wavefunction on the Lorentz group becomes localized at the identity (in the limit).

We  illustrate this explicitly in the 1+1-dimensional case. Consider the  representation~(\ref{final_1+1_k-Poinc_representation}): if it is restricted to act on functions which are localized around $\xi \sim 0$, we can expand all the functions of $\xi$ on the right-hand side around $\xi=0$, and, at first order in $\xi$, the representation looks like:
\bea\label{FirstOrder_kappa_poinc_repr}
a^0  &=&  \ii\lambda \left(\frac 1 2 + q \frac{\partial}{\partial q} \right)  + \ii\lambda  \left(  \frac 1 2 + \xi  \, \frac{\partial}{\partial \xi}\right) +\mathcal{O}(\xi^2) \nonumber\\
 a^1 &=& q + \frac{\ii\lambda}2 \xi  + \mathcal{O}(\xi^2)  \,.
\eea
This reveals  the underlying structure: on wavefunctions sufficiently localized around $\xi=0$, the representation looks like two copies of the $\kappa$-Poincar\'e representation~(\ref{Rep_kappa-Minkowski}), one acting on $q$ and one on $\xi$ (the only difference being that the $\xi$ part of $a^1$ is multiplied by $\ii \lambda /2$, which is irrelevant in our discussion). We are interested in defining a sequence of wavefunctions that localize at $\xi =0$, maintaning the freedom in the choice of the $q$-dependence. The form~(\ref{FirstOrder_kappa_poinc_repr}) suggests to take non-entangled states:
\begin{equation}
\psi_{\sigma,\xi_0}(q,\xi)= f (q) Q_{\sigma,\xi_0} (\xi) \,,
\end{equation} 
Where $Q_{\sigma,\xi_0}$ is a log-Gaussian of similar to~(\ref{LogGaussians}):
\begin{equation}\label{LogGaussian_kappa_poinc}
 Q_{\sigma,\xi_0} (\xi)= \frac{e^{-\frac{\sigma^2}{16}}}{\sqrt{\sqrt{2\pi} \xi_0  \sigma}}  e^{- \left(\frac{\log(\xi^2) - \log (\xi_0^2)}{2 \sigma}\right)^2} \,,
\end{equation}
which is a function which attributes to $\xi^n$ a zero expectation value for $n$ positive and odd, and $e^{\frac{1}{8} n (n+2) \sigma ^2}$ for $n$ positive and even.
\begin{figure}[htb]\center
\includegraphics[width=0.7 \textwidth]{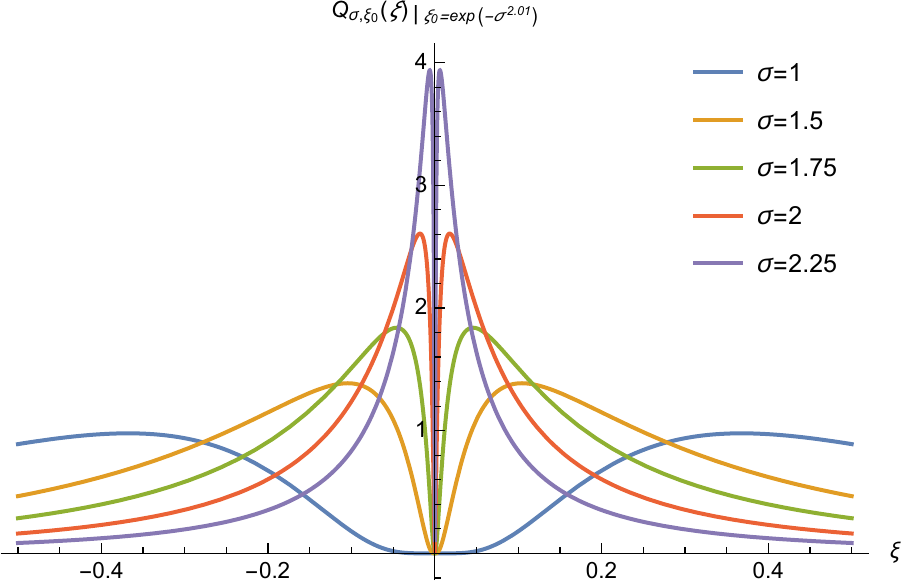}
\caption{\sl The $\sigma \to \infty $ limit or $Q_{\sigma,\xi_0}(\xi)$ when $\xi = e^{-\sigma^{(2+\epsilon)}}$, for $\epsilon = 0.01$.}
\end{figure}

All the expectation values of $(a^\mu)^n$ tend to  
\begin{equation}
\langle \psi_{\sigma,\xi_0} | (a^\mu)^n | \psi_{\sigma,\xi_0} \rangle  \xrightarrow[\xi_0 \to 0 , \sigma \to \infty]{} 
\langle f | (x^\mu)^n | f \rangle  = \int \dd q \bar f(q) (x^\mu)^n f(q) \,,
\end{equation}
where $x^1 = q$ and $x^0 =\ii\lambda \left(\frac 1 2 + q \frac{\partial}{\partial q} \right)$ is the familiar $\kappa$-Poincar\'e representation, and the limits $\xi_0 \to 0$, $\sigma \to \infty$ are taken in such a way that $e^{c \, \sigma^2} \xi_0 \to 0$ for all $c>0$\footnote{As before, we could take $\xi_0 = e^{-  \sigma^{(2 + \epsilon)}}$ and get everything we want from the $\sigma \to \infty$ limit.}.

This is the fundamental content of the statement that $\kappa$-Minkowski is the homogeneous space of $\kappa$-Poincar\'e: we can reproduce any vector $f$ in $L^2(\mathbbm{R}_x)$ taking the limit of the product of $f$ with the log-Gaussian~(\ref{LogGaussian_kappa_poinc}), and all expectation values of powers of translation operators will coincide with the expectation values of the corresponding powers of $x^\mu$ operators on the vector $f$. We reproduce all we know of 
 $\kappa$-Minkowski taking particular states on $\kappa$-Poincar\'e and `silencing' the boost part localizing around $\xi=0$.
 
\section{Observers and Reference Frames \label{se:observers}}

We are representing the algebra~(\ref{commrel}) as generators of operators on the Hilbert space of functions of position. This algebra and its states represent the position in $\kappa$-Minkowski. We have to specify however the \emph{observer} making the observations, and we have been implicitly considering an observer located at the origin. In order to change observer, usually a Poincar\'e transformation is performed. But in our case the symmetry is the quantum $\kappa$-Poincar\'e. Accordingly it will be impossible to locate the position of the transformed observer, since translations do not commute. In the spirit of this paper we will consider the algebra generated by the $a$'s and $\Lambda$'s, and associate to a translated and Lorentz transformed observers a state of this algebra. We first consider the observer located at the origin, which is reached via the identity transformation.

\subsection{The identity transformation state \label{se:identity}}

Looking at the commutation relations~\eqref{kappaPoincareGroup} it is possible to define a state  $\ket{o}_{\mathcal P}$ of $\mathcal P_\kappa$ with the property:
 \begin{equation}
{}_{\mathcal P}\!\bra{o} f(a,\Lambda) | o \rangle_{\mathcal P} = \varepsilon(f) \,,
\end{equation}
where $f(a,\Lambda)$ is a generic element of the $\kappa$-Poincar\'e algebra (\emph{i.e.} a generic noncommutative function of translations and Lorentz transformation matrices), and $\varepsilon$ is the counit of the $\kappa$-Poincar\'e algebra defined in~\eqref{counitantipode}.
In other words the state returns the value of the function on the identity transformation.

We interpret this state in the enlarged algebra as describing the Poincar\'e transformation between two coincident observers, \emph{i.e.} between an observer and a second one located at the origin of the coordinate system of the first observer.
It is not difficult to see, looking at~\eqref{kappaPoincareGroup} that the state is  such that all combined uncertainties vanish. Coincident observers are therefore a well-defined concept in $\kappa$-Minkowski spacetime.

 Note also that all the $\Lambda$'s commute among themselves, and will therefore have common eigenvectors. It is clear from this that the localizability uncertainties have to do with translations, not Lorentz transformations. 

This state can easily be obtained as limit of vectors in the Hilbert space. It suffices to take a succession of functions which converge to a $\delta$ as far as $a^\mu$ and the diagonal elements of $\Lambda^\mu{}_\nu$ are concerned, and to zero for the off-diagonal elements of the $\Lambda$'s.

\subsection{Physical interpretation}

We propose an interpretation for the operators $x^\mu$ we have been using all along, and the operators $x'^\mu$ that appear in Eq.~(\ref{TransformedObserver}): they are the coordinate systems associated to two inertial observers, say, Alice and Bob, which are translated and in relative motion with respect to each other. A spacetime event (\emph{i.e.} the clicking of a particle detector) seen by Alice will be described by the expectation value of its coordinates $\langle x^\mu \rangle$, their variance $\langle (x^\mu - \langle x^\mu \rangle)^2 \rangle$, which measures how localized it is, the skewness  $\langle (x^\mu - \langle x^\mu \rangle)^3 \rangle$ measuring how asymmetric it is around the expectation value, and all higher moments  $\langle (x^\mu - \langle x^\mu \rangle)^n \rangle$  which describe in increasingly finer details the distribution of probability of where the event can be localized. The same event, seen by Bob, will be described by a tower of moments of the transformed coordinate operators: $\langle (x'^\mu - \langle x'^\mu \rangle)^n \rangle$, which are in general different from Alice's, unless the transformation that connects Alice and Bob is the identity described in Sect.~\ref{se:identity}.

What does it mean to take expectation values of the operators $x'^\mu$ and their powers? $x'^\mu$ belongs to the tensor-product algebra $\mathcal{P}_\kappa \otimes \mathcal{M}_\kappa$. We can obtain a representation for this algebra taking the direct sum of the representation~(\ref{3+1kappaPoincareRepresentation}) of $\mathcal{P}_\kappa$ with the representation~(\ref{Rep_kappa-Minkowski}) of $\mathcal{M}_\kappa$.
Clearly the $x^\mu$ algebra (Alice's coordinates) is lifted to elements of the kind $\mathbb 1\otimes \mathcal M_\kappa$, where the identity of $\mathcal P_\kappa$ is given by $\Lambda^\mu{}_\nu=\delta^\mu{}_\nu$, $a^\mu=0$. The  representation of $\mathcal{P}_\kappa \otimes \mathcal{M}_\kappa$ will act on the Hilbert space $\mathcal H_{\mathcal P} \times L^2(\mathbbm{R}^3_x) \sim L^2(SO(3,1)\times\mathbbm{R}^3_q \times \mathbbm{R}^3_x) $, in the following way:
\be
\begin{aligned}
x'^\mu f(\omega,q,x) =&  \ii\lambda \, \Lambda^\mu{}_\nu  (\omega) \left( \delta^\nu{}_0 \, q^i \frac{\partial f(\omega,q,x)}{\partial q^i} + \delta^\nu{}_i \, q^i \, f(\omega,q,x) \right)
\nonumber\\
& + \ii\lambda  \delta^\rho{}_0 \, \left( \frac 3 2 \, \phi(q,\omega)+ q^i \frac{\partial \phi(q,\omega)}{\partial q^i} \right)  + \delta^\mu{}_i \, q^i \, f(\omega,q,x)
\nonumber\\
& - \frac { \ii\lambda} 2 \left[ \left( \Lambda^\mu{}_\sigma \delta^\sigma{}_0 - \delta^\mu{}_0 \right) \Lambda^\rho{}_\nu + \left( \Lambda^\sigma{}_\nu \delta^0{}_\sigma - \delta^0{}_\nu \right) \eta^{\mu\rho} \right] \Lambda^\nu{}_\alpha  \frac{\partial f(\omega,q,x)}{\partial \omega^\mu{}_\alpha} 
\nonumber\\
& -\frac { \ii\lambda} 2  f(\omega,q,x)  \frac{\partial}{\partial \Lambda^\mu{}_\nu}\left[ \left( \Lambda^\mu{}_\sigma \delta^\sigma{}_0 - \delta^\mu{}_0 \right) \Lambda^\rho{}_\nu + \left( \Lambda^\sigma{}_\nu \delta^0{}_\sigma - \delta^0{}_\nu \right) \eta^{\mu\rho} \right] \,.
\end{aligned}
\ee
In the 1+1 dimensional case, we have a more intelligible expression for our representation:
\bea
x'^0 f(\xi,q^1,x^1) &=&    \ii\lambda  \, \cosh \xi \left( \frac 1 2 f+ x^1 \frac{\partial f}{\partial x^1} \right) + \sinh \xi \, x^1  \, f+  \ii\lambda  \left(\frac 1 2 f+ q^1 \frac{\partial f}{\partial q^1} \right)
\nonumber\\&&+  \ii\lambda \left(  \frac 1 2 \cosh \xi \, f + \sinh \xi \, \frac{\partial f}{\partial \xi}\right) \,,
\nonumber\\
x'^1 f(\xi,q^1,x^1) &=& \ii\lambda  \, \sinh \xi \left( \frac 1 2 f+ x^1 \frac{\partial f}{\partial x^1} \right) + \cosh \xi \, x^1  \, f+
  q^1 \, f
 \nonumber\\
  &&+  \ii\lambda  \left( \frac 1 2 \sinh \xi \, f+  \left( \cosh \xi  - 1 \right) \, \frac{\partial f}{\partial \xi}   \right)
 \,.
\eea
 Our Hilbert space will admit  non-entangled states, \emph{i.e.} objects of the kind:
 \be\label{StatePoincareTransformation}
 \ket{g,\psi}=\ket{g}\otimes\ket{\psi}
 \ee
 with $\ket{g}\in\mathcal H_{\mathcal P} = L^2[SO(3,1)]\times\mathbbm{R}^3_q$ and $\ket{\psi}\in L^2(\mathbb  R^3)$.
It represents the state of the coordinates $x'^\mu$ of a Poincar\'e-transformed observer. If we want to calculate the expectation values of the coordinates of the transformed observer we have to do the following:
\begin{equation}
\langle x'^\mu \rangle = \langle g | \otimes \langle \psi | \left(  \Lambda^\mu{}_\nu \otimes x^\nu + a^\mu \otimes 1 \right) | g \rangle \otimes | \psi \rangle
=
 \langle g | \Lambda^\mu{}_\nu | g \rangle
 \langle \psi | x^\nu | \psi \rangle + \langle g | a^\mu | g \rangle \,,
\end{equation}
(we used the normalization condition $\langle \psi | \psi \rangle = 1$). Similarly, one can calculate all the higher momenta of the coordinates as
\begin{equation}
\langle x'^{\mu_1} \dots x'^{\mu_n} \rangle = \langle g | \otimes \langle \psi | \left( x'^{\mu_1} \dots x'^{\mu_n}  \right) | g \rangle \otimes | \psi \rangle \,.
\end{equation}

\subsection{Transforming the states}
We will now derive some general results regarding the properties of these transformed states, which do not depend on a representation except for assuming the existence of the identity state.

\subsubsection{Poincar\'e-transforming the origin state}

Consider the following state which Poincar\'e-tranforms the origin:
\be
\ket{g,0}=\ket{g}\otimes\ket o
\ee
If we want to know what the \emph{Poincar\'e-transformed observer} measures with the coordinates centered on her reference frame, we have to use the operators $x'^\mu  =  \Lambda^\mu{}_\nu \otimes x^\nu + a^\mu \otimes 1$ which act on $L^2(\mathbb{R}^3_x) \times \mathcal H_{\mathcal P}$. Their expectation values on our transformed state are:
\begin{equation}
\langle x'^\mu \rangle =
\langle g | \otimes \langle o | x'^\mu | g \rangle \otimes | o \rangle  =  \langle g | \Lambda^\mu{}_\nu | g \rangle \langle o | x^\nu | o \rangle + \langle g | a^\mu | g \rangle \langle o | o \rangle \,, 
\end{equation}
the state $|o\rangle$ is normalized so $\langle o | o \rangle = 1$, and moreover the expectation value of $x^\mu$ on $|o\rangle$ is, as we have shown
before, zero. We get:
 \begin{equation}
\langle x'^\mu \rangle = \langle g | a^\mu | g \rangle \,, 
\end{equation}
the expectation value of the transformed coordinates is completely determined by the expectation value of the translation operators on the chosen $\kappa$-Poincar\'e state.
 This is natural, the different observers are comparing positions, not directions.
Now consider, more in general, an arbitrary monomial in the transformed coordinates: $x'^{\mu_1}x'^{\mu_2} \dots x'^{\mu_n} $. Its expectation value on $|g \rangle \otimes |o\rangle$ is:
\begin{equation}
\begin{aligned}
\langle  x'^{\mu_1} \dots x'^{\mu_n} \rangle =& \langle g | \otimes \langle o | (a^{\mu_1} \otimes 1 + \Lambda^{\mu_1}{}_{\nu_1} \otimes x^{\nu_1})  \dots (a^{\mu_n} \otimes 1 + \Lambda^{\mu_n}{}_{\nu_n} \otimes x^{\mu_n}) | g \rangle \otimes |o\rangle 
\\
=& \langle g | a^{\mu_1} \dots a^{\mu_n}  | g \rangle \langle o |o\rangle
+
\langle g | \mathcal O^{\mu_1 \dots \mu_n}_\nu (a,\Lambda)  | g \rangle \langle o | x^\nu  |o\rangle + \dots
\\
&
+
\langle g | \mathcal O^{\mu_1 \dots \mu_n}_{\nu_1 \nu_2} (a,\Lambda)  | g \rangle \langle o | x^{\nu_1} x^{\nu_2} |o\rangle
+
\langle g | \mathcal O^{\mu_1 \dots \mu_n}_{\nu_1 \dots \nu_n} (a,\Lambda)  | g \rangle \langle o | x^{\nu_1} \dots x^{\nu_n}  |o\rangle \,,
\end{aligned}
\end{equation}
and since we showed that  $|o\rangle$ is such that $\langle o | x^{\nu_1} \dots x^{\nu_n}  |o\rangle = 0$ $\forall n$,
\begin{equation}
\langle  x'^{\mu_1} \dots x'^{\mu_n} \rangle = 
\langle g | a^{\mu_1} \dots a^{\mu_n}  | g \rangle \langle o |o\rangle =  \langle g| a^{\mu_1} \dots a^{\mu_n} | g \rangle \,.
\end{equation}
Therefore, Poincar\'e transforming the origin state $|o\rangle$ by a state with wavefunction $| g \rangle$ in the representation of  the $\kappa$-Poincar\'e algebra $a^\mu$, $\Lambda^\mu{}_\nu$, the resulting  state will assign, to all polynomials in  the transformed coordinates $x'^\mu = a^\mu \otimes 1 + \Lambda^\mu{}_\nu \otimes x^\nu $, the same expectation value as what assigned by $|g \rangle$ to the corresponding polynomials in $a^\mu$. In other words, the state of $x'^\mu$ is identical to the state of $a^\mu$. So, for example, all uncertainty in the transformed coodinates $\Delta x'^\mu$ is introduced by the uncertainty in the state of the translation operator,  $\Delta a^\mu$.  Let us stress again the fact that, although the new observer is measuring these expectations value, since the $a^\mu$ close a noncommutative algebra, we cannot know, with absolute precision is time and direction, \emph{where} the new observer is, unless she has just time translated the origin, i.e. $\ket g= \ket{o_{a^0}}_{\mathcal P}$.

\subsubsection{Poincar\'e-transforming an arbitrary state with the identity transformation}

A second useful result we  present now is the effect of the identity transformation on an arbitrary state of the $\kappa$-Minkowski coordinates.
 Start from an arbitrary element of the Hilbert space of our representation of the $\kappa$-Minkowski algebra, $| \psi \rangle \in L^2(\mathbbm{R}^3_x)$.
We transform the state as in~(\ref{StatePoincareTransformation}), but using the identity state $\ket{o}_{\mathcal P}$ in place of the generic $|g\rangle$.
On the transformed state $\ket{o}_{\mathcal P} \otimes | \psi \rangle$, all of the expectation values of the polynomials in the transformed coordinates $x'^\mu$  take the form:
\begin{equation}
\begin{aligned}
\langle  x'^{\mu_1} \dots x'^{\mu_n} \rangle =&{}_{\mathcal P}\!\bra{o} \otimes \langle \psi | (a^{\mu_1} \otimes 1 + \Lambda^{\mu_1}{}_{\nu_1} \otimes x^{\nu_1})  \dots (a^{\mu_n} \otimes 1 + \Lambda^{\mu_n}{}_{\nu_n} \otimes x^{\mu_n}) \ket{o}_{\mathcal P} \otimes |\psi \rangle 
\\
=&{}_{\mathcal P}\!\bra{o} a^{\mu_1} \dots a^{\mu_n}  \ket{o}_{\mathcal P} \langle \psi |\psi \rangle
+
{}_{\mathcal P}\!\bra{o}\mathcal O^{\mu_1 \dots \mu_n}_\nu (a,\Lambda)  \ket{o}_{\mathcal P} \langle \psi | x^\nu  |\psi \rangle
\\
&+
{}_{\mathcal P}\!\bra{o} \mathcal O^{\mu_1 \dots \mu_n}_{\nu_1 \nu_2} (a,\Lambda)  \ket{o}_{\mathcal P} \langle \psi | x^{\nu_1} x^{\nu_2} |\psi \rangle
\\
&+ \dots
+
{}_{\mathcal P}\!\bra{o} \mathcal O^{\mu_1 \dots \mu_n}_{\nu_1 \dots \nu_n} (a,\Lambda)  \ket{o}_{\mathcal P} \langle \psi | x^{\nu_1} \dots x^{\nu_n}  |\psi \rangle 
\\
=& \epsilon(a^{\mu_1} \dots a^{\mu_n})\langle \psi |\psi \rangle
+
\epsilon [O^{\mu_1 \dots \mu_n}_\nu (a,\Lambda) ] \langle \psi | x^\nu  |\psi \rangle
+
\epsilon[\mathcal O^{\mu_1 \dots \mu_n}_{\nu_1 \nu_2}] \langle \psi \rangle x^{\nu_1} x^{\nu_2} |\psi \rangle
\\
&+ \dots
+
\epsilon[\mathcal O^{\mu_1 \dots \mu_n}_{\nu_1 \dots \nu_n} (a,\Lambda)] \langle \psi | x^{\nu_1} \dots x^{\nu_n}  |\psi \rangle \,,
\end{aligned}
\end{equation}
now, the algebra elements $O^{\mu_1 \dots \mu_n}_{\nu_1 \dots \nu_m} (a,\Lambda)$ are monomials in $a^\mu$, $\Lambda^\mu{}_\nu$, without a particular ordering. However, we know that the $m$-th element contains $m$ Lorentz matrix generators and $n-m$ translation generators. Using the homomorphism property of the counit map $\epsilon$, and the fact that  $\epsilon(a^\mu)=0$, $\epsilon(\Lambda^\mu{}_\nu)=\delta^\mu{}_\nu$, we can prove that
\begin{equation}
\epsilon[ O^{\mu_1 \dots \mu_n}_{\nu_1 \dots \nu_m} (a,\Lambda) ] = 0 \, ~~ \text{unless } m=n \,, 
\end{equation}
and
\begin{equation}
\epsilon[ O^{\mu_1 \dots \mu_n}_{\nu_1 \dots \nu_n} (a,\Lambda) ] =
\delta^{\mu_1}{}_{\nu_1} \dots  \delta^{\mu_n}_{\nu_n} \,,
\end{equation}
we conclude that
\begin{equation}
\begin{aligned}
{}_{\mathcal P}\!\bra{o}\otimes \langle \psi |  x'^{\mu_1} \dots x'^{\mu_n}  \ket{o}_{\mathcal P} \otimes |\psi \rangle  = 
 \langle \psi | x^{\mu_1} \dots x^{\mu_n}  |\psi \rangle \,,
\end{aligned}
\end{equation}
\emph{i.e.}, the identity transformation does not change any expectation value - the original observer (who uses the coordinate operators $x^\mu$ and the Hilbert space $L^2(\mathbbm{R}^3_x)$), and the transformed one (using the coordinates operators $x'^\mu$ and the Hilbert space $\mathbbm{H}_{\mathcal P} \otimes L^2(\mathbbm{R}^3_x)$), agree on all measurements if the state of $\mathcal{H}_{\mathcal P} $ that defines the transformation is~$\ket{o}_{\mathcal P}$.

\subsubsection{$\kappa$-Poincar\'e and coordinate uncertainty}

Consider a generic transformation of a generic state: $|\psi \rangle \to |g \rangle \otimes |\psi \rangle $. We want to study the relationship between the uncertainty in the transformed coordinates $\Delta x'^\mu$ and the one of the original ones $ \Delta x^\mu$.

First, the simplest example: a pure translation $x'^\mu = 1 \otimes x^\mu + a^\mu \otimes 1$. Calculating the variance of $x^\mu$:
\begin{equation}
\begin{aligned}
\Delta ( x'^\mu )^2 =&  \langle (x'^\mu)^2 \rangle -  \langle x'^\mu \rangle^2 
=
 \langle (x^\mu)^2 + (a^\mu)^2 + x^\mu a^\mu + a^\mu x^\mu\rangle -  \langle x^\mu \rangle^2 -  \langle a^\mu \rangle^2 -  2 \langle x^\mu \rangle  \langle a^\mu \rangle
 \\
 =& \Delta ( x^\mu )^2 + \Delta ( a^\mu )^2 + 2 \, \text{cov}(x^\mu , a^\mu) \,.
\end{aligned}
\end{equation}
The covariance between $a^\mu$ and $x^\mu$ is zero, because they belong to different sides of the tensor product:
\begin{equation}
\begin{aligned}
2 \, \text{cov}(x^\mu , a^\mu) &= 
  \langle g | \otimes \langle \psi | (  x^\mu a^\mu + a^\mu x^\mu ) |g \rangle \otimes |\psi \rangle  -  2 \langle \psi | x^\mu | \psi  \rangle  \langle g | a^\mu | g \rangle
  \\
&= \langle \psi |x^\mu|\psi \rangle  \langle g | a^\mu |g \rangle  + \langle g | a^\mu |g \rangle \langle \psi |x^\mu |\psi \rangle  -  2 \langle \psi | x^\mu | \psi  \rangle  \langle g | a^\mu | g \rangle = 0 \,,
\end{aligned}
\end{equation}
we conclude that:
\begin{equation}\label{UncertaintyIncreaseTranslations}
\begin{aligned}
\Delta ( x'^\mu )^2 = \Delta ( x^\mu )^2 + \Delta ( a^\mu )^2  \geq  \Delta ( x^\mu )^2 \,,
\end{aligned}
\end{equation}
\emph{i.e.} a translation can only increase the uncertainty of the coordinates. One is simply adding uncorrelated variables, and their uncertainties get square-summed.\footnote{Notice that this conclusion is a consequence of the fact that we assumed that transformed states are product states $| g \rangle \otimes | \psi \rangle$. If we allowed for entanglement between the transformation part $| g \rangle$ and the state $| \psi \rangle$ describing the event in the initial reference frame, we would have opened the possibility of reducing the uncertainty of $x^\mu$ with a translation. This, however, conflicts with the basic physical intuition that the relationship between inertial observers should be independent of the state of the system that the observers are studying.}

Performing a translation results in an increase of the uncertainty in the coordinates, unless the translation parameter has zero uncertainty. This happens only in the cases of the identity transformation \emph{or of a purely-temporal translation}, which can have zero uncertainty in all of the $a^\mu$'s, in analogy with the discussion in the introduction. We have the nice result that the uncertainty do not depend on time translations.

Consider a state which looks uncertain to the observer Alice located at the origin. One could think that there would be another observer, Bob, translated with respect to Alice, such that this same state is perfectly localised for him. 
One could naively think to start (in 1+1D) from the state $\psi(x^1)$  for $x^1$, and then make a translation with wavefunction $\psi(-q^1)$ where $\psi$ is the same function. One would think that the  translated state is localized at the origin. Relation~\eqref{UncertaintyIncreaseTranslations} shows that this is impossible. Calculating the expectation value of $(x'^1)^n = (x^1 + a^1)^n$ a Newton binomial sum of this kind is obtained:
\bea
\langle (x^1 + a^1)^n \rangle
&=&
\sum_{m=0}^n \left( \begin{array}{c} n \\ m \end{array} \right)  \langle \psi(x^1) | (x^1)^{n-m} |\psi(x^1)\rangle \langle \psi(-q) | (a^1)^m |\psi(-q)\rangle =
\nonumber\\&=& 
\sum_{m=0}^n \left( \begin{array}{c} n \\ m \end{array} \right)  \langle \psi | (x^1)^{n-m} |\psi\rangle  \langle \psi | (-x^1)^{m} |\psi\rangle 
\eea
The above expression can never be zero. For example, for $n=2$:
\be
\langle (x^1 + a^1)^2 \rangle = \langle (x^1)^2 \rangle + 2 \langle x^1 a^1 \rangle  + \langle( a^1)^2 \rangle
= 2 \langle (x^1)^2 \rangle - 2 \langle x^1\rangle^2 = 2 \Delta(x^1)^2
\ee
The variance doubles, it does not go to zero!

The process of translating a state and then ``undo'' it with a change of observer does not lead to an identification of states. Of course the symmetry between Alice and Bob is preserved, each has a set of states which is isomorphic, but the quantum nature of the transformation implies that this set of states is not transformed into each other by a translation.

Now let's consider general $\kappa$-Poincar\'e transformations, for example  the transformation of the spatial coordinate in 1+1 dimensions:
\begin{equation}
x'^1 = \cosh \xi \otimes x^1 + \sinh \xi \otimes x^0 + a^1 \otimes 1 \,.
\end{equation}
calculating the difference between its variance and the variance of $x^1$:
\begin{equation}
\begin{aligned}
&\Delta(x'^1)^2 = \Delta(x^1)^2  +
\Delta(a^1)^2 + \langle x^1 \rangle^2 \Delta(\cosh \xi)^2 + \langle x^0 \rangle ^2 \Delta(\sinh \xi)^2
\\
&  + \langle \sinh \xi \rangle^2 \Delta(x^0)^2 + \Delta(\sinh \xi)^2 \Delta(x^0)^2 + \langle \cosh \xi \rangle^2 \Delta(x^1)^2 + \Delta(\cosh \xi)^2 \Delta(x^1)^2
 \\
& +2 \, \text{cov}(x^1, x^0) \langle \cosh \xi \rangle \langle \sinh \xi \rangle+ 2 \, \text{cov}(a^1,\sinh \xi) \langle x^0 \rangle + 2 \, \text{cov}(a^1, \cosh \xi) \langle x^1 \rangle 
\\
&+ 2 \, \text{cov}(\cosh \xi, \sinh \xi) (\text{cov}(x^0, x^1) + \langle x^0 \rangle \langle x^1 \rangle) - \Delta(x^1)^2 \,,
\end{aligned}
\end{equation}
the above expression can be rewritten as 
\begin{equation}
\begin{aligned}
\Delta(x'^1)^2 =& \Delta(x^1)^2 +
\langle \sinh^2 \xi \rangle  \left( \Delta(x^0)^2 + \Delta(x^1)^2\right)  
\\
&+  \Delta[\cosh \xi]^2 \langle x^1 \rangle^2 + \Delta[\sinh \xi]^2 \langle x^0 \rangle^2 + 2 \text{cov}(\cosh \xi ,\sinh \xi) \langle x^0 \rangle \langle x^1 \rangle 
\\
& + \Delta[a^1]^2 + 2 \text{cov}(\cosh \xi,a^1) \langle x^1 \rangle  
 + 2  \text{cov}(\sinh \xi,a^1) \langle x^0 \rangle
 \\
 &
+ 2 \langle \cosh \xi \sinh \xi \rangle \text{cov}(x^0,x^1)  \,,
\end{aligned}
\end{equation}
the second and third lines above can be rewritten as the squared uncertainty of the operator $a^1+ \sinh \xi \, \langle x^0 \rangle + \cosh \xi \, \langle x^1 \rangle$, which is positive, and we get:
\begin{equation}\label{ChangeUncertaintyPoincare}
\begin{aligned}
\Delta(x'^1)^2 - \Delta(x^1)^2 =& \Delta[a^1+ \sinh \xi \, \langle x^0 \rangle + \cosh \xi \, \langle x^1 \rangle ]^2 
\\
&+\langle \sinh^2 \xi \rangle \left( \Delta(x^0)^2 + \Delta(x^1)^2\right)
+ 2 \langle \cosh \xi \sinh \xi \rangle \text{cov}(x^0,x^1)  \,.
\end{aligned}
\end{equation}
Now, assume that  $\langle x^0 \rangle =\langle x^1 \rangle $ so that the first term reduces to the uncertainty of $a^1$. Moreover, we rewrite the covariance of $x^0$ and $x^1$ as $2 \text{cov}(x^0,x^1) = \Delta(x^0+x^1)^2  -\Delta(x^0)^2 - \Delta(x^1)^2 $:
\begin{equation}
\begin{aligned}
\Delta(x'^1)^2 - \Delta(x^1)^2 =& \Delta(a^1)^2 +
\left(\langle \sinh^2 \xi \rangle - \langle \cosh \xi \sinh \rangle \right) \left( \Delta(x^0)^2 + \Delta(x^1)^2\right)
\\
&+ \langle \cosh \xi \sinh \xi \rangle \Delta(x^0+x^1)^2   \,.
\end{aligned}
\end{equation}
It is easy to prove that:
\begin{equation}
\langle \sinh^2 \xi \rangle +  \langle \cosh \xi \sinh \xi \rangle 
=
{\textstyle \frac 1 2} \left( \langle e^{2 \xi} \rangle - 1  \right) \,,
\end{equation}
so that:
\begin{equation}
\begin{aligned}
\Delta(x'^1)^2 - \Delta(x^1)^2 =& \Delta(a^1)^2 +
{\textstyle \frac 1 2} \left( \langle e^{2 \xi} \rangle - 1  \right)  \left( \Delta(x^0)^2 + \Delta(x^1)^2\right)
\\
&+ \langle \cosh \xi \sinh \xi \rangle \Delta(x^0+x^1)^2   \,.
\end{aligned}
\end{equation}
One linear combination of $x^0$ and $x^1$ can always be made arbitrarily localized, so we can make $\Delta(x^0+x^1)^2$ arbitrarily small. The same of course holds for $\Delta(a^1)^2$, without putting any constraint on the other quantities except the uncertainty of $\xi$, which however doesn't limit much our ability to manipulate the state in order to adjust the values of $\langle e^{2 \xi} \rangle$ and $ \langle \cosh \xi \sinh \xi \rangle$. It doesn't take long to convince oneself that we can concoct a state such that $\langle e^{2 \xi} \rangle < 1$ (\emph{e.g.} it is sufficient that the wavefunction over $\xi$ be supported on the $\xi <0$ region), and $ \langle \cosh \xi \sinh \xi \rangle$ is $\mathcal{O}(1)$. Then the expression above will be dominated by 
${\textstyle \frac 1 2} \left( \langle e^{2 \xi} \rangle - 1  \right)  \left( \Delta(x^0)^2 + \Delta(x^1)^2\right)$ which is negative.

We proved that the variances of $x^\mu$ can only increase after a pure translation, but, under particular circumstances, they can decrease after a Poincar\'e transformations. In particular, states with zero expectation value of $x^\mu$ such that the uncertainty of $(x^0 + x^1)$ is sufficiently small, can reduce their uncertainty if we perform a $\kappa$-Poincar\'e transformation with sufficiently localized translation and a Lorentz transformation such that $\langle e^{2 \xi} \rangle < 1$ and $ \langle \cosh \xi \sinh \xi \rangle = \mathcal{O}(1)$. We postpone to further work the study of the physical consequences of this observation.

\section{Conclusions and outlook}

In this paper we discussed a way to look at the $\kappa$-Minkowski quantum space with the tools of the algebra of operators and the theory of measurement initially developed for ordinary quantum mechanics. This enables a coherent way to look at states, localization, and transformations. The picture of quantum $\kappa$-Minkowski spacetime which emerges is, in our opinion, quite fascinating. There are no absolutely localized points, but it is nevertheless possible to find states which approximately localise. The role of Fourier transformation from position to momentum is here played by Mellin transforms which connect time with (radial) position. We also laid out the foundations of a discussion of the deformed transformations of this space. This is an aspect which will deserve further scrutiny for a complete understanding of transformation theory. In this paper we presented a series of basic results valid in 3+1 dimensions, and discussed in quantitative details the 1+1-dimensional case. Generalizing all of our results to the 3+1-dimensional case seems technically more complicated, but there do not seem to be any conceptual obstacle.
   A possible future development could be addressing the fact that we used a particular representation of the operators, while other are possible. It should be investigated if the alternatives are, at least qualitatively, similar.

Finally the next challenge: we considered a regime which is not very natural in physics, namely we considered the effects of a quantum spacetime for which the noncommutativity parameter of space, $\lambda$ is nonzero, while $\hbar$ can be ignored. Bringing $\hbar$ back into the picture would require us to consider momenta  (either in the form of wave modes in a field-theoretical setting, or as quantity of motion of particles). The space of momenta in $\kappa$-Minkowski is curved~\cite{KappaDeSitterMomentumSpace,MicheleCurvedMomentum,kappaRelativeLocality,
MercatiSergolaPauliJordan}, and this has led to introduce the principle of relative locality~\cite{PRErelativelocality,relativelocality,kappaRelativeLocality}. The relationship between the relaxations of locality we found in the present paper and those introduced by relative locality is an interesting open issue, worth exploring.

\subsection*{Acknowledgements}
The authors acknowledge the COST action QSPACE, and in particular the Short Term Scientific Missions which enabled two visits of TP in Napoli. FL and MM acknowledge the support of  the INFN Iniziativa Specifica GeoSymQFT; FL the Spanish MINECO under project MDM-2014-0369 of ICCUB (Unidad de Excelencia `Maria de Maeztu').
FM has received funding from the European Union's research and innovation programme under a Marie Sk\l{}odowska-Curie grant through the  INdAM-COFUND-2012 programme of the Italian Institue of High Mathematics (INdAM).
We thank Florio M.~Ciaglia, Max Kurkov, Marco Laudato, Patrizia Vitale and Jean-Christophe Wallet for discussions.

\end{document}